\documentclass[aps,prd,preprint,groupedaddress,nofootinbib,showkeys,
sort&compress]{revtex4-1}

\usepackage{amsmath,amssymb}
\usepackage{booktabs,longtable,array,tabularx}
\usepackage{float}
\usepackage{microtype}
\usepackage{etoolbox}
\usepackage{xcolor}
\colorlet{red}{black}
\usepackage[colorlinks=true,allcolors=blue]{hyperref}

\newcommand{\CP}{\mathit{CP}}

\newcommand{\RS}{\mathcal{S}}
\newcommand{\Pzero}{\mathcal{P}^{0}}
\newcommand{\Vzero}{\mathcal{V}^{0}}
\newcommand{\Bzero}{\mathcal{B}^{0}}
\newcolumntype{Y}[1]{>{\raggedright\arraybackslash}p{#1}}
\AtBeginEnvironment{thebibliography}{\sloppy\hbadness=2000\relax}

\begin{document}

\title{New \texorpdfstring{$\CP$}{CP} sum rules from exact
\texorpdfstring{$U$}{U}-spin degeneracy}

\author{Chao-Qiang Geng$^1$}\email{cqgeng@ucas.ac.cn}
\author{Chia-Wei Liu$^1$}\email{chiaweiliu@ucas.ac.cn}
\author{Sheng-Lin Liu$^{1,2,3}$}\email{liushenglin22@mails.ucas.ac.cn}
\author{Bing-Bo-Min Shang$^{1,2,3}$}\email{shangbingbomin25@mails.ucas.ac.cn}
\affiliation{$^1$School of Fundamental Physics and Mathematical Sciences, Hangzhou Institute for Advanced Study, UCAS, Hangzhou 310024, China}
\affiliation{$^2$Institute of Theoretical Physics,~UCAS, Beijing 100190, China}
\affiliation{$^3$University of Chinese Academy of Sciences, 100190 Beijing, China}

% Corresponding-author information should be supplied before journal submission.

\date{\today}

\begin{abstract}
We propose a new class of $\CP$ sum rules in the $U$-spin limit, based on
the observation that the weak phase in the Standard Model can be removed in
the degenerate limit $m_d=m_s$.  Unlike previous $U$-spin analyses formulated
at the amplitude level, our sum rules are expressed directly in terms of decay
widths.  They apply to matched partial waves, spin observables, and angular
moments and include nonpairwise cases beyond the usual single-insertion
$d\leftrightarrow s$ relation.
\end{abstract}

%\keywords{$\CP$ violation, $U$-spin  flavor symmetry, sum rules}

\maketitle

In the Standard Model, flavor change in charged-current quark interactions
is governed by the Cabibbo--Kobayashi--Maskawa (CKM) matrix~\cite{Kobayashi:1973fv, Chau:1984fp}:
\begin{eqnarray}
 V_{\mathrm{CKM}}& = &
 \begin{pmatrix}
 1&0&0\\
 0&c_{23}&s_{23}\\
 0&-s_{23}&c_{23}
 \end{pmatrix} 
 \begin{pmatrix}
 c_{13}&0&s_{13}e^{-i\delta_{13}}\\
 0&1&0\\
 -s_{13}e^{i\delta_{13}}&0&c_{13}
 \end{pmatrix}
 \begin{pmatrix}
 c_{12}&s_{12}&0\\
 -s_{12}&c_{12}&0\\
 0&0&1
 \end{pmatrix},    
 \label{eq:ckm-matrix}
\end{eqnarray}
where $c_{ij}=\cos\theta_{ij}$ and $s_{ij}=\sin\theta_{ij}$.  For three
generations, the mass-dependent measure of $\CP$ violation is given by the
Jarlskog invariant~\cite{Jarlskog:1985ht},
\begin{equation}
 \begin{aligned}
 \mathcal I_{\CP}
  &=J_{\mathrm{CKM}}
 (m_t^2-m_c^2)(m_t^2-m_u^2)(m_c^2-m_u^2)
 (m_b^2-m_s^2)(m_b^2-m_d^2)(m_s^2-m_d^2),\\[-1mm]
 J_{\mathrm{CKM}}
 &=c_{12}c_{23}c_{13}^{2}s_{12}s_{23}s_{13}\sin\delta_{13}.
 \end{aligned}
 \label{eq:jarlskog-invariant}
\end{equation}
At $m_d=m_s$, $\mathcal I_{\CP}$ vanishes.  Assuming 
$\delta_{13}$ 
 is
the only $\CP$-odd source, the enlarged $U(2)$ basis freedom and quark
rephasings can make $V_{\mathrm{CKM}}$ real~\cite{Bernabeu:1986fc,Ecker:1987qp}.

However, one need not rotate away the $\CP$ phase.  For example,
QCD-factorization and perturbative-QCD calculations yield a nonzero
$\CP$-odd rate difference for
$B_d^0\to\pi^-K^+$~\cite{Beneke:2001ev,Keum:2000wi}.  Formally continuing
such a fixed-flavor-basis calculation to $m_d=m_s$ need not make this
channel-specific quantity vanish.  This apparent contradiction
arises because $(\pi^-,K^-)$ and $(\pi^+,K^+)$ span degenerate $U$-spin
doublets in this limit.  A decay rate assigned to the single basis state
$\pi^-K^+$ is therefore not basis invariant and is not by itself a physical
observable in the exact-degeneracy theory.  The physical observable must
instead be summed over the corresponding complete degenerate subspace.
Throughout, we define the $\CP$-odd rate difference by
$\Gamma_{\CP}(i\to f)\equiv\Gamma(i\to f)-\Gamma(\bar i\to\bar f)$,
where barred labels denote the $\CP$-conjugate states.  The integrated
$\CP$-odd widths obey
\begin{equation} 
 \boxed{\displaystyle
 \sum_{i\in I,\,f\in F}\Gamma_{\CP}(i\to f)=0.}
 \label{eq:operator-sum}
\end{equation}
Here $I$ and $F$ sum over the degenerate states that are indistinguishable at $m_d = m_s$.
Partial-wave widths obey analogous sum rules, modified by the
appropriate identical-particle symmetry factors so that the full-width $\CP$ sum rule in
Eq.~\eqref{eq:operator-sum} is restored after integration.

Equation~\eqref{eq:operator-sum} can also be understood directly from the
completeness relation.  With $P_X=\sum_{x\in X}|x\rangle\langle x|$, the complete rate is
\begin{eqnarray} 
 \sum _{i \in I ,f \in F  } 
 \Gamma ( i\to f ; T) 
 &\propto &
 \sum_{i\in I,\,f\in F}
 |\langle f|T|i\rangle|^2 , 
 \label{eq:trace}
\end{eqnarray}
where $T$ is the transition operator.  Since a unitary basis change $R$
within a complete multiplet maps one complete orthonormal basis into another,
its projector is invariant:
\begin{equation}
 R  P_X R  ^\dagger =\sum_{x\in X}|R  x\rangle\langle R  x|
     = P_X, 
 \label{eq:rearrangement}
\end{equation}
with $X= I,F$.  
Equivalently,
\begin{equation}\label{eq:rate-basis-invariance}
 \sum _{i \in I ,f \in F  } 
\Gamma ( i\to f ; T) 
= \sum _{i \in I ,f \in F  } 
\Gamma ( i\to f ; R T R^\dagger ).
\end{equation} 
At $m_d=m_s$, one can take $R$ to implement the rephasings
$q\to e^{i\varphi_q}q$ for $q=u,c,t,b$ and
$(d,s)^T\to U_{ds}(d,s)^T$, with
\begin{equation}
	\begin{aligned}
		&  \varphi_b=\varphi_c=\varphi_t=\delta_{13},
		\qquad \varphi_u=0,\\[-1mm]
		& U_{ds} =
		\begin{pmatrix}
			c_{12}&-s_{12}\\
			s_{12}& c_{12}
		\end{pmatrix}
		\begin{pmatrix}
			\cos\vartheta&\sin\vartheta\\
			-\sin\vartheta e^{i\delta_{13}}
			&\cos\vartheta e^{i\delta_{13}}
		\end{pmatrix},
	\end{aligned}
	\label{eq:real-basis-rotation}
\end{equation}
where $\vartheta$ is arbitrary.
This transformation removes $\delta_{13}$ from
Eq.~\eqref{eq:ckm-matrix} and shows that the right-hand side of
Eq.~\eqref{eq:rate-basis-invariance} contains no $\CP$ violation.

% MAIN-TEXT TABLE INSERTION POINT

% BEGIN GENERATED MAIN-TEXT TABLES
% This region is generated from the verified nontrivial tables in
% ../../all_results.tex by ../../tmp/build_prd_appendix.pl.
\setlength{\LTcapwidth}{0.96\columnwidth}

\begin{table}[t]
\caption{Nontrivial   $U$-spin sum rules for the  $B\to PV$ and $B\to PP$ decays.  Here
$[i\to f]$ denotes $\Gamma_{\CP}(i\to f)$.}
\label{tab:B-meson-nontrivial}
\centering
\begin{tabular}{@{}Y{0.96\linewidth}@{}}
\hline
\hline 
$\CP$-odd rate-difference sum rule\\
\hline
\(\displaystyle
[B^-\to\pi^-\rho^0]
+[B^-\to\pi^-\omega]
+[B^-\to\pi^-\phi]{}+ 
[B^-\to K^-\rho^0]
+[B^-\to K^-\omega]
+[B^-\to K^-\phi]\)\\[2pt]\hline
\(\displaystyle
[B^-\to\pi^-\pi^0]+[B^-\to\pi^-\eta]+[B^-\to\pi^-\eta']{}+ 
[B^-\to K^-\pi^0]+[B^-\to K^-\eta]+[B^-\to K^-\eta']\)\\[2pt]\hline
\(\displaystyle
[B^-\to\pi^0\rho^-]
\!+\![B^-\to\pi^0K^{*-}]
\!+\![B^-\to\eta\rho^-]
\!+\![B^-\to\eta K^{*-}]
\!+\![B^-\to\eta'\rho^-]
\!+\![B^-\to\eta'K^{*-}]\)\\[2pt]\hline
\(\displaystyle
[\bar B^0\to\bar K^0\rho^0]
+[\bar B^0\to\bar K^0\omega]
+[\bar B^0\to\bar K^0\phi]{}+ 
[\bar B_s^0\to K^0\rho^0]
+[\bar B_s^0\to K^0\omega]
+[\bar B_s^0\to K^0\phi]\)\\[2pt]\hline
\(\displaystyle
[\bar B^0\to\pi^0\bar K^{*0}]
+[\bar B^0\to\eta\bar K^{*0}]
+[\bar B^0\to\eta'\bar K^{*0}]{}+ 
[\bar B_s^0\to\pi^0K^{*0}]
+[\bar B_s^0\to\eta K^{*0}]
+[\bar B_s^0\to\eta'K^{*0}]\)\\[2pt]\hline
\(\displaystyle
[\bar B^0\to\pi^0\rho^0]
+[\bar B^0\to\eta\rho^0]
+[\bar B^0\to\eta'\rho^0]
+[\bar B^0\to\pi^0\omega]{}+ 
[\bar B^0\to\eta\omega]
+[\bar B^0\to\eta'\omega]
+[\bar B^0\to\pi^0\phi]
+[\bar B^0\to\eta\phi]{}+ 
[\bar B^0\to\eta'\phi]
+[\bar B_s^0\to\pi^0\rho^0]
+[\bar B_s^0\to\eta\rho^0]
+[\bar B_s^0\to\eta'\rho^0]
+[\bar B_s^0\to\pi^0\omega]{}+ 
[\bar B_s^0\to\eta\omega]
\!+\![\bar B_s^0\to\eta'\omega]
\!+\![\bar B_s^0\to\pi^0\phi]
\!+\![\bar B_s^0\to\eta\phi]
\!+\![\bar B_s^0\to\eta'\phi]\)\\[2pt]\hline
\(\displaystyle
[\bar B^0\to\pi^0\pi^0]+[\bar B^0\to\pi^0\eta]+[\bar B^0\to\pi^0\eta']+[\bar B^0\to\eta\eta]{}+ 
[\bar B^0\to\eta\eta']+[\bar B^0\to\eta'\eta']+[\bar B_s^0\to\pi^0\pi^0]+[\bar B_s^0\to\pi^0\eta]{}+ 
[\bar B_s^0\to\pi^0\eta']+[\bar B_s^0\to\eta\eta]+[\bar B_s^0\to\eta\eta']+[\bar B_s^0\to\eta'\eta']\)\\[2pt]\hline
\(\displaystyle
[\bar B^0\to\bar K^0\pi^0]+[\bar B^0\to\bar K^0\eta]+[\bar B^0\to\bar K^0\eta']{} 
+[\bar B_s^0\to K^0\pi^0]+[\bar B_s^0\to K^0\eta]+[\bar B_s^0\to K^0\eta']\)\\[2pt]
\hline
\hline 
\end{tabular}
\end{table}
% END GENERATED MAIN-TEXT TABLES

For comparison, consider the standard $U$-spin reflection relation with a
single weak-Hamiltonian insertion.  Applications to $\CP$ asymmetries in
beauty-meson decays have been developed in
Refs.~\cite{Deshpande:1994ii,He:1998rq,Fleischer:1999pa,Gronau:2000zy,
 Dariescu:2002hw,Lipkin:2005pb,Nagashima:2007qn,Imbeault:2011jz,
 He:2013vta,Bhattacharya:2013cvn,Gronau:2013mda,Grossman:2013lya,
 Xu:2013rua,Bhattacharya:2015uua,He:2017fln,Nir:2022bbh,
 Fleischer:2022rkm}.  Related charm analyses are given in
Refs.~\cite{Grossman:2006jg,Pirtskhalava:2011va,Grossman:2012ry,
 Muller:2015rna,Schacht:2022kuj,Bause:2022jes,Iguro:2024uuw}.  Baryon
applications appear in Refs.~\cite{He:2015fwa,Gronau:2015gha,He:2015fsa,
 Grossman:2018ptn,Wang:2019dls,Roy:2019cky,Roy:2020nyx,Wang:2024rwf,
 Zhang:2025jnw,Roy:2025nao,He:2025msg,Wen:2025ibn}.  General $U$-spin
sum-rule systematics are discussed in
Refs.~\cite{Gavrilova:2022hbx,Gavrilova:2024npn}.  Let
$R=R_2(\pi)$, which sends $d\to s$ and $s\to-d$.  Then
\begin{equation}
\Gamma_{\CP}(i\to f)+\Gamma_{\CP}(Ri\to Rf)=0.
 \label{eq:pairwise-reflection-integrated}
\end{equation}
If each external hadron appearing in $i$ and $f$ is a $U$-spin eigenstate,
Eq.~\eqref{eq:operator-sum} becomes
\begin{equation}
\!\!\! \sum_{[(i,f)]\in(I\times F)/R}
 \left[\Gamma_{\CP}(i\to f)
 +\Gamma_{\CP}(Ri\to Rf)\right]=0.
 \label{eq:pairwise-orbit-sum}
\end{equation}
Each two-element orbit obeys Eq.~\eqref{eq:pairwise-reflection-integrated};
thus Eq.~\eqref{eq:operator-sum} is simply their sum.
We stress that our Eq.~\eqref{eq:operator-sum} holds to all orders in the weak interactions, whereas Eq.~\eqref{eq:pairwise-reflection-integrated} holds only at leading order. 

The genuinely nontrivial consequences of Eq.~\eqref{eq:operator-sum}
therefore arise when at least one state entering $i$ or $f$ is not a
$U$-spin eigenstate, as occurs for $(\pi^0,\eta_8)$,
$(\rho^0,\omega,\phi)$, and $(\Lambda,\Sigma^0)$.
The states in each group are indistinguishable in the limit $m_u=m_d=m_s$ with exact isospin symmetry.
For $(\pi^0,\eta_8)$, the physical $\eta$  is a mixed state of $\eta_8$ and $\eta_0$. To extract the effects, we can enlarge the summation from $(\pi ^0 , \eta _8)$ to $(\pi ^0 , \eta , \eta ')$\footnote{
In the full light-quark $SU(3)$ limit, $m_u=m_d=m_s$, the octet and singlet states do not mix, so one may identify $\eta=\eta_8$ and $\eta'=\eta_0$~\cite{Feldmann:1998vh}. Therefore, the sum rules, in the strict sense, only need to involve the $(\pi^0,\eta)$ states in the degenerate limit.}.
  In the physical basis,
these states are coherent mixtures of different $U$-spin components.  The
reflection therefore mixes several   channels, so there is no unique
partner in Eq.~\eqref{eq:pairwise-reflection-integrated} and the
complete-multiplet sum must be kept. For instance, one such nontrivial sum rule is
\begin{equation}
 \sum_{P^-=\pi^-,K^-}\ \sum_{V^0=\rho^0,\omega , \phi }
\Gamma _{\CP}(B^-\to P^-V^0)=0.
\label{eq:pv-complete-sum}
\end{equation}
The charge-conjugate, phase-space-stripped counterpart of this relation was
derived by the amplitude method of Ref.~\cite{Grossman:2013lya} in the $SU(3)$
limit, under ideal $\omega$--$\phi$ mixing and with a single weak-Hamiltonian
insertion.  The
complete-width sum in Eq.~\eqref{eq:pv-complete-sum}, derived from the projector
trace above, is instead basis independent and requires no assumption about
$\omega$--$\phi$ mixing. Using the unrounded
branching fractions and direct $\CP$ asymmetries underlying
Ref.~\cite{ParticleDataGroup:2026rpp}, the convention above gives
$\tau_{B^-}\Gamma_{\CP}(B^-\to f)=2\mathcal B_f^{\rm av}A_{\CP}(f)$.
Treating the inputs as uncorrelated and adding the quoted uncertainties in
quadrature, we obtain
\begin{equation}
 \tau_{B^-}\sum_{f }\Gamma_{\CP}(B^-\to f)
 =(+0.64\pm0.90)\times10^{-6}, 
 \label{eq:charged-b-pv-data-test}
\end{equation} 
from the experimental data~\cite{ParticleDataGroup:2026rpp}, which is consistent with the sum rule.
 Here $\tau_{B^-}$ is the lifetime of $B^-$. 

Table~\ref{tab:B-meson-nontrivial} lists the charged
$B^-\to P^-V^0$ relation in Eq.~\eqref{eq:pv-complete-sum} as its first row
and the charged $B^-\to P^-P^0$ relation as its second row, followed by four
additional nontrivial $B\to PV$ relations and two additional nontrivial
$B\to PP$ relations.  Some of these
sums have been derived by the 
amplitude analyses~\cite{Grossman:2013lya}.
The second-row relation is also testable with current data~\cite{ParticleDataGroup:2026rpp} 
\begin{equation}
 \begin{aligned}
 \!\! \tau_{B^-}\sum_{f }\Gamma_{\CP}(B^-\to f) 
 =(-1.43\pm1.93)\times10^{-6},
 \end{aligned}
 \label{eq:charged-b-data-test}
\end{equation}
with $f$ summed over the final states of the second row in the table. 
   Beyond these two charged-$B$ sums, no other
nontrivial table entry currently has all the required branching-fraction and
direct-$\CP$-asymmetry inputs.   

{\color{red}
A phenomenologically useful but trivial illustration is provided by
the charged $D\to PP$ mode. At one local $\Delta C=1$ insertion in the exact $U$-spin limit,
Eq.~\eqref{eq:operator-sum} reduces to
$
\Gamma_{\CP}(D^0\to K^+K^-)
+\Gamma_{\CP}(D^0\to\pi^+\pi^-)=0
$, with the CF and DCS contributions vanishing separately. LHCb finds
$a_{\CP}^{\rm dir}(K^+K^-)=(7.7\pm5.7)\times10^{-4}$ and
$a_{\CP}^{\rm dir}(\pi^+\pi^-)=(23.2\pm6.1)\times10^{-4}$~\cite{LHCb:2022lry}.  Using
$\tau_{D^0}\Gamma_{\CP}(f)=2\mathcal B_f^{\rm av}a_f^{\rm dir}$ and the
HFLAV averaged data~\cite{HeavyFlavorAveragingGroupHFLAV:2024ctg}, we obtain
\begin{equation}
 2\left(
 \mathcal B_{KK}^{\rm av}a_{KK}^{\rm dir}
 +\mathcal B_{\pi\pi}^{\rm av}a_{\pi\pi}^{\rm dir}
 \right)
 =(13.24\pm6.34)\times10^{-6},
 \label{eq:d0-scs-data-test}
\end{equation}
where the correlation between the two measured asymmetries has been retained.
Thus the rate-weighted $U$-spin relation differs from zero by $2.1\sigma$.
On the other hand, the theoretical estimates of the same left-hand side all have magnitudes below
$2\times10^{-6}$~\cite{Li:2012cfa,Khodjamirian:2017zdu,Cheng:2019ggx,
	Bediaga:2022sxw,Pich:2023kim}, and are therefore much closer to zero than
the experimental value in Eq.~\eqref{eq:d0-scs-data-test}.
}

We collect the remaining nontrivial two-body meson identities in
Appendix~\ref{sec:appendix-channel-catalogs},
Eqs.~\eqref{eq:app-1.1}--\eqref{eq:app-1.9}.
Equations~\eqref{eq:app-1.1}--\eqref{eq:app-1.3} give the
$D,D_s\to PV$ relations,
Eqs.~\eqref{eq:app-1.4}--\eqref{eq:app-1.7} complete the
$D,D_s\to PP,VV$ sectors, and
Eqs.~\eqref{eq:app-1.8} and \eqref{eq:app-1.9} provide the two neutral
$B,B_s\to VV$ relations complementary to the charged relation discussed
below.  The charged-$D$ neutral-vector sum in
Eq.~\eqref{eq:app-1.2} also has a phase-space-stripped amplitude-level
counterpart~\cite{Grossman:2013lya}.

The construction extends directly to multibody decays, for which
such relations are difficult to deduce with a conventional amplitude
analysis.  Appendix~\ref{sec:appendix-channel-catalogs} organizes the complete
multibody catalog in the same notation: Eqs.~\eqref{eq:app-2.1}--
\eqref{eq:app-2.11} give the three-body charmed-baryon identities,
Eqs.~\eqref{eq:app-3.1}--\eqref{eq:app-4.28} give the three-body
bottom-baryon identities, and Eqs.~\eqref{eq:app-5.1}--\eqref{eq:app-6.106} give the four-body bottom-baryon identities.  All of
the displayed relations use partial widths integrated over common matched
regions.

For two-body channels, the differential identity underlying
Eq.~\eqref{eq:operator-sum} may be applied on each matched phase-space element.
Consequently, the integrated relation applies to any common region $\Omega$
and separately to each partial wave.  For a spin-0 meson
decaying into two vector mesons, define the coefficients~\cite{Datta:2007qb}
\begin{equation}
 \begin{aligned}
 K_a={}&\bigl(|A_0|^2,|A_\parallel|^2,|A_\perp|^2,
 \operatorname{Re}A_\parallel A_0^*,\operatorname{Im}A_\perp A_0^*,
 \operatorname{Im}A_\perp A_\parallel^*\bigr),\\
 \eta_a={}&(1,1,1,1,-1,-1),\qquad
 \Gamma_{\CP}^{a}=K_a-\eta_a\overline K_a.
 \end{aligned}
 \label{eq:transversity}
\end{equation}
Here $A_0$, $A_\perp$, and $A_\parallel$ are the partial-wave amplitudes, and
$\Gamma_{\CP}^{a}$ is the corresponding $\CP$-odd transversity coefficient.
In the exact-symmetry limit it follows that
\begin{equation}
 \sum_{ V^+  , 
V^0  
} 
  \Gamma^a _{\CP} (B^+  \to  V^ +   V^0 )=0,
 \label{eq:vector-vector-complete-sum}
\end{equation}
with $V^+\in\{K^{*+},\rho^+\}$ and
$V^0\in\{\rho^0,\omega,\phi\}$.  These partial-wave sum rules can be tested in the near future. 

We conclude with baryons.  In a two-body decay of a spin-$1/2$ particle into
a spin-$1/2$ particle and a spinless particle,
the Lee--Yang parameters are~\cite{Lee:1957qs,Donoghue:1986hh}
\begin{equation}
\color{red}
 \alpha=\frac{2\operatorname{Re}(A_S^*A_P)}
 {|A_S|^2+|A_P|^2},\quad
 \beta=\frac{2\operatorname{Im}(A_S^*A_P)}
 {|A_S|^2+|A_P|^2},\quad
 \gamma=\frac{|A_S|^2-|A_P|^2}
 {|A_S|^2+|A_P|^2}.
\label{eq:lee-yang}
\end{equation}
\begingroup
\color{red}
Here $A_S$ and $A_P$ denote the $S$- and $P$-wave decay amplitudes,
respectively; the symbol $P$ without an amplitude label is reserved for a
pseudoscalar meson.
\endgroup
In a common
particle--antiparticle angular convention, $\CP$ conservation implies
$\bar\alpha=-\alpha$, $\bar\beta=-\beta$, and $\bar\gamma=\gamma$.
\begingroup
\color{red}
For a channel $i\to f$, let $\Gamma_f$ and $\bar\Gamma_{\bar f}$ denote
the partial widths of the particle and $\CP$-conjugate decays, respectively,
and let $\xi_f$ and $\bar\xi_{\bar f}$, for
$\xi\in\{\alpha,\beta,\gamma\}$, denote the corresponding Lee--Yang
coefficients measured in the same angular convention.  For
$\xi\in\{\Gamma,\alpha,\beta,\gamma\}$, we define
\begin{equation}
 \xi_{\CP}(i\to f)\equiv
 \begin{cases}
  \Gamma_f-\bar\Gamma_{\bar f}, & \xi=\Gamma,\\[1mm]
  \Gamma_f\xi_f-\eta_\xi\bar\Gamma_{\bar f}\bar\xi_{\bar f},
  & \xi\in\{\alpha,\beta,\gamma\},
 \end{cases}
\label{eq:lee-yang-cp}
\end{equation}
with
\begin{equation*}
(\eta_\alpha,\eta_\beta,\eta_\gamma)=(-1,-1,+1).
\end{equation*}
In the $\CP$ limit, $\bar\xi_{\bar f}=\eta_\xi\xi_f$.
Thus $\Gamma_{\CP}$ is a rate difference, whereas
$\alpha_{\CP}$, $\beta_{\CP}$, and $\gamma_{\CP}$ are rate-weighted angular
moments, not partial widths.
\endgroup
An example of a nontrivial sum rule is
\begingroup
\color{red}
\begin{eqnarray}
&&  
	\xi_{\CP} ( 
\Xi _b ^ -  \to \Lambda \pi  ^- ) 
+ 	\xi_{\CP} ( 
\Xi _b ^ -  \to \Sigma^ 0   \pi ^- )      
\label{eq:bottom-baryon-example-sum}
\\&& 
+  	\xi_{\CP} ( 
\Xi _b ^ -  \to \Lambda K ^- ) 
+ 	\xi_{\CP} ( 
\Xi _b ^ -  \to \Sigma^ 0   K ^- )   = 0. \nonumber 
\end{eqnarray} 
\endgroup
This example is the shortest member of the baryon catalog.  The remaining
two-body baryon identities are collected in
Eqs.~\eqref{eq:app-7.1}--\eqref{eq:app-12.3}, following the multibody relations
described above without a change of notation.

\section*{Acknowledgments} 
This work was supported in part by the National Natural Science Foundation of China
(NSFC) under Grant Nos. 12547104 and 12575096.

\appendix
% BEGIN INLINE APPENDIX
% Channel content generated from CP_Uspin_PRD_supplement_equations.tex.
% Narrow-column layout reflowed by ../../tmp/reflow_appendix_equations.pl.
\section{Catalog of \texorpdfstring{$\CP$}{CP} sum rules for two- and multibody decays}

\label{sec:appendix-channel-catalogs}

We collect the remaining two-body and complete multibody channel catalogs with
the neutral-state shorthands
\begin{equation*}
  \begin{gathered}
    \Pzero\equiv\{\pi^0,\,\eta,\,\eta'\},\qquad
    \Vzero\equiv\{\rho^0,\,\omega,\,\phi\},\\[-1pt]
    \Bzero\equiv\{\Lambda,\,\Sigma^0\}.
  \end{gathered}
\end{equation*}
The equations below use the full decay widths with the shorthand of 
\begin{equation}
  \RS(I;F_1,\ldots,F_n)
  \equiv
  \sum_{\substack{i\in I\\ f_a\in F_a}}
  \frac{\Gamma_{\CP}(i\to f_1\cdots f_n)}
       {N_{\mathrm{perm}}(f_1,\ldots,f_n)}.
  \tag{A0}\label{eq:app-0}
\end{equation} 
where $N_{\mathrm{perm}}$ counts the ordered assignments representing the same
physical final state and removes permutation double counting.

{\footnotesize
\setlength{\parskip}{0pt}
\setlength{\jot}{1pt}
\setlength{\abovedisplayskip}{-4pt}
\setlength{\belowdisplayskip}{-4pt}
\setlength{\abovedisplayshortskip}{-4pt}
\setlength{\belowdisplayshortskip}{-4pt}
\begin{align}
\RS\bigl(\{D^0\};\,\Vzero;\,\Pzero\bigr)&={}\,0 \tag{A1.1}\label{eq:app-1.1}
\end{align}

\begin{align}
\RS\bigl(\{D^+\};\,\Vzero;\,\{\pi^+\}\bigr){}+{}\RS\bigl(\{D_s^+\};\,\Vzero;\,\{K^+\}\bigr)&={}\,0 \tag{A1.2}\label{eq:app-1.2}
\end{align}

\begin{align}
\RS\bigl(\{D^+\};\,\{\rho^+\};\,\Pzero\bigr){}+{}\RS\bigl(\{D_s^+\};\,\{K^{*+}\};\,\Pzero\bigr)&={}\,0 \tag{A1.3}\label{eq:app-1.3}
\end{align}

\begingroup
\color{red}
\begin{align}
\RS\bigl(\{D^0\};\,\Pzero;\,\Pzero\bigr)&={}\,0 \tag{A1.4}\label{eq:app-1.4}
\end{align}

\begin{align}
\RS\bigl(\{D^+\};\,\{\pi^+\};\,\Pzero\bigr){}+{}\RS\bigl(\{D_s^+\};\,\{K^+\};\,\Pzero\bigr)&={}\,0 \tag{A1.5}\label{eq:app-1.5}
\end{align}

\begin{align}
\RS\bigl(\{D^0\};\,\Vzero;\,\Vzero\bigr)&={}\,0 \tag{A1.6}\label{eq:app-1.6}
\end{align}

\begin{align}
\RS\bigl(\{D^+\};\,\{\rho^+\};\,\Vzero\bigr){}+{}\RS\bigl(\{D_s^+\};\,\{K^{*+}\};\,\Vzero\bigr)&={}\,0 \tag{A1.7}\label{eq:app-1.7}
\end{align}

\begin{align}
\RS\bigl(\{\bar B^0,\,\bar B_s^0\};\,\{K^{*0},\,\bar K^{*0}\};\,\Vzero\bigr)&={}\,0 \tag{A1.8}\label{eq:app-1.8}
\end{align}

\begin{align}
\RS\bigl(\{\bar B^0,\,\bar B_s^0\};\,\Vzero;\,\Vzero\bigr)&={}\,0 \tag{A1.9}\label{eq:app-1.9}
\end{align}
\endgroup

\begin{align}
\RS\bigl(\{\Lambda_c^+\};\,\{p\};\,\Pzero;\,\Pzero\bigr){}+{}\RS\bigl(\{\Xi_c^+\};\,\{\Sigma^+\};\,\Pzero;\,\Pzero\bigr)&={}\,0 \tag{A2.1}\label{eq:app-2.1}
\end{align}

\begin{align}
\RS\bigl(\{\Lambda_c^+\};\,\{n\};\,\{\pi^+\};\,\Pzero\bigr){}+{}\RS\bigl(\{\Xi_c^+\};\,\{\Xi^0\};\,\Pzero;\,\{K^+\}\bigr)&={}\,0 \tag{A2.2}\label{eq:app-2.2}
\end{align}

\begin{align}
\RS\bigl(\{\Lambda_c^+\};\,\{\Sigma^+\};\,\{K^0\};\,\Pzero\bigr){}+{}\RS\bigl(\{\Xi_c^+\};\,\{p\};\,\{\bar K^0\};\,\Pzero\bigr)&={}\,0 \tag{A2.3}\label{eq:app-2.3}
\end{align}

\begin{align}
\RS\bigl(\{\Lambda_c^+\};\,\Bzero;\,\{\pi^+\};\,\{K^0\}\bigr){}+{}\RS\bigl(\{\Xi_c^+\};\,\Bzero;\,\{K^+\};\,\{\bar K^0\}\bigr)&={}\,0 \tag{A2.4}\label{eq:app-2.4}
\end{align}

\begin{align}
\RS\bigl(\{\Lambda_c^+\};\,\Bzero;\,\{K^+\};\,\Pzero\bigr){}+{}\RS\bigl(\{\Xi_c^+\};\,\Bzero;\,\{\pi^+\};\,\Pzero\bigr)&={}\,0 \tag{A2.5}\label{eq:app-2.5}
\end{align}

\begin{align}
\RS\bigl(\{\Xi_c^0\};\,\{\Sigma^+\};\,\{\pi^-\};\,\Pzero\bigr){}+{}\RS\bigl(\{\Xi_c^0\};\,\{p\};\,\{K^-\};\,\Pzero\bigr)&={}\,0 \tag{A2.6}\label{eq:app-2.6}
\end{align}

\begin{align}
\RS\bigl(\{\Xi_c^0\};\,\{\Xi^0\};\,\{K^0\};\,\Pzero\bigr){}+{}\RS\bigl(\{\Xi_c^0\};\,\{n\};\,\{\bar K^0\};\,\Pzero\bigr)&={}\,0 \tag{A2.7}\label{eq:app-2.7}
\end{align}

\begin{align}
\RS\bigl(\{\Xi_c^0\};\,\Bzero;\,\{K^+\};\,\{K^-\}\bigr){}+{}\RS\bigl(\{\Xi_c^0\};\,\Bzero;\,\{\pi^+\};\,\{\pi^-\}\bigr)&={}\,0 \tag{A2.8}\label{eq:app-2.8}
\end{align}

\begin{align}
\RS\bigl(\{\Xi_c^0\};\,\{\Sigma^-\};\,\{\pi^+\};\,\Pzero\bigr){}+{}\RS\bigl(\{\Xi_c^0\};\,\{\Xi^-\};\,\Pzero;\,\{K^+\}\bigr)&={}\,0 \tag{A2.9}\label{eq:app-2.9}
\end{align}

\begin{align}
\RS\bigl(\{\Xi_c^0\};\,\Bzero;\,\Pzero;\,\Pzero\bigr)&={}\,0 \tag{A2.10}\label{eq:app-2.10}
\end{align}

\begin{align}
\RS\bigl(\{\Xi_c^0\};\,\Bzero;\,\{K^0\};\,\{\bar K^0\}\bigr)&={}\,0 \tag{A2.11}\label{eq:app-2.11}
\end{align}

\begin{align}
\RS\bigl(\{\Xi_b^-\};\,\Bzero;\,\Pzero;\,\{K^-,\,\pi^-\}\bigr)&={}\,0 \tag{A3.1}\label{eq:app-3.1}
\end{align}

\begin{align}
\RS\bigl(\{\Xi_b^-\};\,\Bzero;\,\{K^-\};\,\{K^0\}\bigr){}+{}\RS\bigl(\{\Xi_b^-\};\,\Bzero;\,\{\pi^-\};\,\{\bar K^0\}\bigr)&={}\,0 \tag{A3.2}\label{eq:app-3.2}
\end{align}

\begin{align}
\RS\bigl(\{\Xi_b^-\};\,\{\Xi^0\};\,\Pzero;\,\{\pi^-\}\bigr){}+{}\RS\bigl(\{\Xi_b^-\};\,\{n\};\,\Pzero;\,\{K^-\}\bigr)&={}\,0 \tag{A3.3}\label{eq:app-3.3}
\end{align}

\begin{align}
\RS\bigl(\{\Xi_b^-\};\,\{\Sigma^-,\,\Xi^-\};\,\Pzero;\,\Pzero\bigr)&={}\,0 \tag{A3.4}\label{eq:app-3.4}
\end{align}

\begin{align}
\RS\bigl(\{\Xi_b^-\};\,\{\Sigma^-\};\,\Pzero;\,\{\bar K^0\}\bigr){}+{}\RS\bigl(\{\Xi_b^-\};\,\{\Xi^-\};\,\Pzero;\,\{K^0\}\bigr)&={}\,0 \tag{A3.5}\label{eq:app-3.5}
\end{align}

\begin{align}
\RS\bigl(\{\Lambda_b^0,\,\Xi_b^0\};\,\Bzero;\,\Pzero;\,\Pzero\bigr)&={}\,0 \tag{A3.6}\label{eq:app-3.6}
\end{align}

\begin{align}
\RS\bigl(\{\Lambda_b^0\};\,\Bzero;\,\Pzero;\,\{K^0\}\bigr){}+{}\RS\bigl(\{\Xi_b^0\};\,\Bzero;\,\Pzero;\,\{\bar K^0\}\bigr)&={}\,0 \tag{A3.7}\label{eq:app-3.7}
\end{align}

\begin{align}
&\phantom{{}+{}}\RS\bigl(\{\Lambda_b^0\};\,\Bzero;\,\{K^-,\,\pi^-\};\,\{K^+\}\bigr){}+{}\RS\bigl(\{\Lambda_b^0\};\,\Bzero;\,\{\pi^-\};\,\{\pi^+\}\bigr) \notag\\
&{}+{}\RS\bigl(\{\Xi_b^0\};\,\Bzero;\,\{K^-,\,\pi^-\};\,\{\pi^+\}\bigr){}+{}\RS\bigl(\{\Xi_b^0\};\,\Bzero;\,\{K^+\};\,\{K^-\}\bigr)={}\,0 \tag{A3.8}\label{eq:app-3.8}
\end{align}

\begin{align}
\RS\bigl(\{\Lambda_b^0,\,\Xi_b^0\};\,\Bzero;\,\{\bar K^0\};\,\{K^0\}\bigr)&={}\,0 \tag{A3.9}\label{eq:app-3.9}
\end{align}

\begin{align}
\RS\bigl(\{\Lambda_b^0\};\,\{n\};\,\Pzero;\,\Pzero\bigr){}+{}\RS\bigl(\{\Xi_b^0\};\,\{\Xi^0\};\,\Pzero;\,\Pzero\bigr)&={}\,0 \tag{A3.10}\label{eq:app-3.10}
\end{align}

\begin{align}
\RS\bigl(\{\Lambda_b^0,\,\Xi_b^0\};\,\{\Xi^0\};\,\Pzero;\,\{K^0\}\bigr){}+{}\RS\bigl(\{\Lambda_b^0,\,\Xi_b^0\};\,\{n\};\,\Pzero;\,\{\bar K^0\}\bigr)&={}\,0 \tag{A3.11}\label{eq:app-3.11}
\end{align}

\begin{align}
&\phantom{{}+{}}\RS\bigl(\{\Lambda_b^0\};\,\{\Sigma^-\};\,\Pzero;\,\{K^+,\,\pi^+\}\bigr){}+{}\RS\bigl(\{\Lambda_b^0\};\,\{\Xi^-\};\,\Pzero;\,\{K^+\}\bigr) \notag\\
&{}+{}\RS\bigl(\{\Xi_b^0\};\,\{\Sigma^-\};\,\Pzero;\,\{\pi^+\}\bigr){}+{}\RS\bigl(\{\Xi_b^0\};\,\{\Xi^-\};\,\Pzero;\,\{K^+,\,\pi^+\}\bigr)={}\,0 \tag{A3.12}\label{eq:app-3.12}
\end{align}

\begin{align}
&\phantom{{}+{}}\RS\bigl(\{\Lambda_b^0\};\,\{\Sigma^+\};\,\Pzero;\,\{\pi^-\}\bigr){}+{}\RS\bigl(\{\Lambda_b^0\};\,\{p\};\,\Pzero;\,\{K^-,\,\pi^-\}\bigr) \notag\\
&{}+{}\RS\bigl(\{\Xi_b^0\};\,\{\Sigma^+\};\,\Pzero;\,\{K^-,\,\pi^-\}\bigr){}+{}\RS\bigl(\{\Xi_b^0\};\,\{p\};\,\Pzero;\,\{K^-\}\bigr)={}\,0 \tag{A3.13}\label{eq:app-3.13}
\end{align}

\begin{align}
\RS\bigl(\{\Xi_b^-\};\,\Bzero;\,\{K^{*-},\,\rho^-\};\,\Pzero\bigr)&={}\,0 \tag{A4.1}\label{eq:app-4.1}
\end{align}

\begin{align}
\RS\bigl(\{\Xi_b^-\};\,\Bzero;\,\Vzero;\,\{K^-,\,\pi^-\}\bigr)&={}\,0 \tag{A4.2}\label{eq:app-4.2}
\end{align}

\begin{align}
\RS\bigl(\{\Xi_b^-\};\,\Bzero;\,\{K^{*0}\};\,\{K^-\}\bigr){}+{}\RS\bigl(\{\Xi_b^-\};\,\Bzero;\,\{\bar K^{*0}\};\,\{\pi^-\}\bigr)&={}\,0 \tag{A4.3}\label{eq:app-4.3}
\end{align}

\begin{align}
\RS\bigl(\{\Xi_b^-\};\,\Bzero;\,\{K^{*-}\};\,\{K^0\}\bigr){}+{}\RS\bigl(\{\Xi_b^-\};\,\Bzero;\,\{\rho^-\};\,\{\bar K^0\}\bigr)&={}\,0 \tag{A4.4}\label{eq:app-4.4}
\end{align}

\begin{align}
\RS\bigl(\{\Xi_b^-\};\,\{\Xi^0\};\,\{\rho^-\};\,\Pzero\bigr){}+{}\RS\bigl(\{\Xi_b^-\};\,\{n\};\,\{K^{*-}\};\,\Pzero\bigr)&={}\,0 \tag{A4.5}\label{eq:app-4.5}
\end{align}

\begin{align}
\RS\bigl(\{\Xi_b^-\};\,\{\Xi^0\};\,\Vzero;\,\{\pi^-\}\bigr){}+{}\RS\bigl(\{\Xi_b^-\};\,\{n\};\,\Vzero;\,\{K^-\}\bigr)&={}\,0 \tag{A4.6}\label{eq:app-4.6}
\end{align}

\begin{align}
\RS\bigl(\{\Xi_b^-\};\,\{\Sigma^-,\,\Xi^-\};\,\Vzero;\,\Pzero\bigr)&={}\,0 \tag{A4.7}\label{eq:app-4.7}
\end{align}

\begin{align}
\RS\bigl(\{\Xi_b^-\};\,\{\Sigma^-\};\,\{\bar K^{*0}\};\,\Pzero\bigr){}+{}\RS\bigl(\{\Xi_b^-\};\,\{\Xi^-\};\,\{K^{*0}\};\,\Pzero\bigr)&={}\,0 \tag{A4.8}\label{eq:app-4.8}
\end{align}

\begin{align}
\RS\bigl(\{\Xi_b^-\};\,\{\Sigma^-\};\,\Vzero;\,\{\bar K^0\}\bigr){}+{}\RS\bigl(\{\Xi_b^-\};\,\{\Xi^-\};\,\Vzero;\,\{K^0\}\bigr)&={}\,0 \tag{A4.9}\label{eq:app-4.9}
\end{align}

\begin{align}
\RS\bigl(\{\Lambda_b^0,\,\Xi_b^0\};\,\Bzero;\,\Vzero;\,\Pzero\bigr)&={}\,0 \tag{A4.10}\label{eq:app-4.10}
\end{align}

\begin{align}
\RS\bigl(\{\Lambda_b^0\};\,\Bzero;\,\{K^{*0}\};\,\Pzero\bigr){}+{}\RS\bigl(\{\Xi_b^0\};\,\Bzero;\,\{\bar K^{*0}\};\,\Pzero\bigr)&={}\,0 \tag{A4.11}\label{eq:app-4.11}
\end{align}

\begingroup
\color{red}
\begin{align}
\RS\bigl(\{\Lambda_b^0,\,\Xi_b^0\};\,\Bzero;\,\{\rho^+\};\,\{\pi^-\}\bigr){}+{}\RS\bigl(\{\Lambda_b^0,\,\Xi_b^0\};\,\Bzero;\,\{K^{*+}\};\,\{K^-\}\bigr)&={}\,0 \tag{A4.12}\label{eq:app-4.12}
\end{align}
\endgroup

\begingroup
\color{red}
\begin{align}
\RS\bigl(\{\Xi_b^0\};\,\Bzero;\,\{\rho^+\};\,\{K^-\}\bigr){}+{}\RS\bigl(\{\Lambda_b^0\};\,\Bzero;\,\{K^{*+}\};\,\{\pi^-\}\bigr)&={}\,0 \tag{A4.13}\label{eq:app-4.13}
\end{align}
\endgroup

\begingroup
\color{red}
\begin{align}
\RS\bigl(\{\Lambda_b^0,\,\Xi_b^0\};\,\Bzero;\,\{\rho^-\};\,\{\pi^+\}\bigr){}+{}\RS\bigl(\{\Lambda_b^0,\,\Xi_b^0\};\,\Bzero;\,\{K^{*-}\};\,\{K^+\}\bigr)&={}\,0 \tag{A4.14}\label{eq:app-4.14}
\end{align}
\endgroup

\begingroup
\color{red}
\begin{align}
\RS\bigl(\{\Xi_b^0\};\,\Bzero;\,\{K^{*-}\};\,\{\pi^+\}\bigr){}+{}\RS\bigl(\{\Lambda_b^0\};\,\Bzero;\,\{\rho^-\};\,\{K^+\}\bigr)&={}\,0 \tag{A4.15}\label{eq:app-4.15}
\end{align}
\endgroup

\begin{align}
\RS\bigl(\{\Lambda_b^0\};\,\Bzero;\,\Vzero;\,\{K^0\}\bigr){}+{}\RS\bigl(\{\Xi_b^0\};\,\Bzero;\,\Vzero;\,\{\bar K^0\}\bigr)&={}\,0 \tag{A4.16}\label{eq:app-4.16}
\end{align}

\begin{align}
\RS\bigl(\{\Lambda_b^0,\,\Xi_b^0\};\,\Bzero;\,\{\bar K^{*0}\};\,\{K^0\}\bigr){}+{}\RS\bigl(\{\Lambda_b^0,\,\Xi_b^0\};\,\Bzero;\,\{K^{*0}\};\,\{\bar K^0\}\bigr)&={}\,0 \tag{A4.17}\label{eq:app-4.17}
\end{align}

\begin{align}
\RS\bigl(\{\Lambda_b^0\};\,\{n\};\,\Vzero;\,\Pzero\bigr){}+{}\RS\bigl(\{\Xi_b^0\};\,\{\Xi^0\};\,\Vzero;\,\Pzero\bigr)&={}\,0 \tag{A4.18}\label{eq:app-4.18}
\end{align}

\begin{align}
\RS\bigl(\{\Lambda_b^0,\,\Xi_b^0\};\,\{\Xi^0\};\,\{K^{*0}\};\,\Pzero\bigr){}+{}\RS\bigl(\{\Lambda_b^0,\,\Xi_b^0\};\,\{n\};\,\{\bar K^{*0}\};\,\Pzero\bigr)&={}\,0 \tag{A4.19}\label{eq:app-4.19}
\end{align}

\begin{align}
\RS\bigl(\{\Lambda_b^0,\,\Xi_b^0\};\,\{\Xi^0\};\,\Vzero;\,\{K^0\}\bigr){}+{}\RS\bigl(\{\Lambda_b^0,\,\Xi_b^0\};\,\{n\};\,\Vzero;\,\{\bar K^0\}\bigr)&={}\,0 \tag{A4.20}\label{eq:app-4.20}
\end{align}

\begingroup
\color{red}
\begin{align}
\RS\bigl(\{\Xi_b^0\};\,\{\Xi^-\};\,\{\rho^+\};\,\Pzero\bigr){}+{}\RS\bigl(\{\Lambda_b^0\};\,\{\Sigma^-\};\,\{K^{*+}\};\,\Pzero\bigr)&={}\,0 \tag{A4.21}\label{eq:app-4.21}
\end{align}
\endgroup

\begingroup
\color{red}
\begin{align}
\RS\bigl(\{\Lambda_b^0,\,\Xi_b^0\};\,\{\Sigma^-\};\,\{\rho^+\};\,\Pzero\bigr){}+{}\RS\bigl(\{\Lambda_b^0,\,\Xi_b^0\};\,\{\Xi^-\};\,\{K^{*+}\};\,\Pzero\bigr)&={}\,0 \tag{A4.22}\label{eq:app-4.22}
\end{align}
\endgroup

\begingroup
\color{red}
\begin{align}
\RS\bigl(\{\Xi_b^0\};\,\{\Xi^-\};\,\Vzero;\,\{\pi^+\}\bigr){}+{}\RS\bigl(\{\Lambda_b^0\};\,\{\Sigma^-\};\,\Vzero;\,\{K^+\}\bigr)&={}\,0 \tag{A4.23}\label{eq:app-4.23}
\end{align}
\endgroup

\begingroup
\color{red}
\begin{align}
\RS\bigl(\{\Lambda_b^0,\,\Xi_b^0\};\,\{\Sigma^-\};\,\Vzero;\,\{\pi^+\}\bigr){}+{}\RS\bigl(\{\Lambda_b^0,\,\Xi_b^0\};\,\{\Xi^-\};\,\Vzero;\,\{K^+\}\bigr)&={}\,0 \tag{A4.24}\label{eq:app-4.24}
\end{align}
\endgroup

\begingroup
\color{red}
\begin{align}
\RS\bigl(\{\Xi_b^0\};\,\{\Sigma^+\};\,\{K^{*-}\};\,\Pzero\bigr){}+{}\RS\bigl(\{\Lambda_b^0\};\,\{p\};\,\{\rho^-\};\,\Pzero\bigr)&={}\,0 \tag{A4.25}\label{eq:app-4.25}
\end{align}
\endgroup

\begingroup
\color{red}
\begin{align}
\RS\bigl(\{\Lambda_b^0,\,\Xi_b^0\};\,\{\Sigma^+\};\,\{\rho^-\};\,\Pzero\bigr){}+{}\RS\bigl(\{\Lambda_b^0,\,\Xi_b^0\};\,\{p\};\,\{K^{*-}\};\,\Pzero\bigr)&={}\,0 \tag{A4.26}\label{eq:app-4.26}
\end{align}
\endgroup

\begingroup
\color{red}
\begin{align}
\RS\bigl(\{\Xi_b^0\};\,\{\Sigma^+\};\,\Vzero;\,\{K^-\}\bigr){}+{}\RS\bigl(\{\Lambda_b^0\};\,\{p\};\,\Vzero;\,\{\pi^-\}\bigr)&={}\,0 \tag{A4.27}\label{eq:app-4.27}
\end{align}
\endgroup

\begingroup
\color{red}
\begin{align}
\RS\bigl(\{\Lambda_b^0,\,\Xi_b^0\};\,\{\Sigma^+\};\,\Vzero;\,\{\pi^-\}\bigr){}+{}\RS\bigl(\{\Lambda_b^0,\,\Xi_b^0\};\,\{p\};\,\Vzero;\,\{K^-\}\bigr)&={}\,0 \tag{A4.28}\label{eq:app-4.28}
\end{align}
\endgroup

\begin{align}
\RS\bigl(\{\Xi_b^-\};\,\Bzero;\,\{K^-,\,\pi^-\};\,\Pzero;\,\Pzero\bigr)&={}\,0 \tag{A5.1}\label{eq:app-5.1}
\end{align}

\begin{align}
\RS\bigl(\{\Xi_b^-\};\,\Bzero;\,\{K^-\};\,\{K^0\};\,\Pzero\bigr){}+{}\RS\bigl(\{\Xi_b^-\};\,\Bzero;\,\{\pi^-\};\,\{\bar K^0\};\,\Pzero\bigr)&={}\,0 \tag{A5.2}\label{eq:app-5.2}
\end{align}

\begin{align}
\RS\bigl(\{\Xi_b^-\};\,\Bzero;\,\{K^-,\,\pi^-\};\,\{K^-\};\,\{K^+\}\bigr){}+{}\RS\bigl(\{\Xi_b^-\};\,\Bzero;\,\{K^-,\,\pi^-\};\,\{\pi^-\};\,\{\pi^+\}\bigr)&={}\,0 \tag{A5.3}\label{eq:app-5.3}
\end{align}

\begin{align}
\RS\bigl(\{\Xi_b^-\};\,\Bzero;\,\{K^-,\,\pi^-\};\,\{\bar K^0\};\,\{K^0\}\bigr)&={}\,0 \tag{A5.4}\label{eq:app-5.4}
\end{align}

\begin{align}
\RS\bigl(\{\Xi_b^-\};\,\{\Xi^0\};\,\{\pi^-\};\,\Pzero;\,\Pzero\bigr){}+{}\RS\bigl(\{\Xi_b^-\};\,\{n\};\,\{K^-\};\,\Pzero;\,\Pzero\bigr)&={}\,0 \tag{A5.5}\label{eq:app-5.5}
\end{align}

\begin{align}
\RS\bigl(\{\Xi_b^-\};\,\{\Xi^0\};\,\{K^-,\,\pi^-\};\,\{K^0\};\,\Pzero\bigr){}+{}\RS\bigl(\{\Xi_b^-\};\,\{n\};\,\{K^-,\,\pi^-\};\,\{\bar K^0\};\,\Pzero\bigr)&={}\,0 \tag{A5.6}\label{eq:app-5.6}
\end{align}

\begin{align}
\RS\bigl(\{\Xi_b^-\};\,\{\Sigma^-,\,\Xi^-\};\,\Pzero;\,\Pzero;\,\Pzero\bigr)&={}\,0 \tag{A5.7}\label{eq:app-5.7}
\end{align}

\begin{align}
\RS\bigl(\{\Xi_b^-\};\,\{\Sigma^-\};\,\{\bar K^0\};\,\Pzero;\,\Pzero\bigr){}+{}\RS\bigl(\{\Xi_b^-\};\,\{\Xi^-\};\,\{K^0\};\,\Pzero;\,\Pzero\bigr)&={}\,0 \tag{A5.8}\label{eq:app-5.8}
\end{align}

\begin{align}
&\phantom{{}+{}}\RS\bigl(\{\Xi_b^-\};\,\{\Sigma^-\};\,\{K^-,\,\pi^-\};\,\{\pi^+\};\,\Pzero\bigr){}+{}\RS\bigl(\{\Xi_b^-\};\,\{\Sigma^-\};\,\{K^-\};\,\{K^+\};\,\Pzero\bigr) \notag\\
&{}+{}\RS\bigl(\{\Xi_b^-\};\,\{\Xi^-\};\,\{K^-,\,\pi^-\};\,\{K^+\};\,\Pzero\bigr){}+{}\RS\bigl(\{\Xi_b^-\};\,\{\Xi^-\};\,\{\pi^+\};\,\{\pi^-\};\,\Pzero\bigr)={}\,0 \tag{A5.9}\label{eq:app-5.9}
\end{align}

\begin{align}
\RS\bigl(\{\Xi_b^-\};\,\{\Sigma^-,\,\Xi^-\};\,\{\bar K^0\};\,\{K^0\};\,\Pzero\bigr)&={}\,0 \tag{A5.10}\label{eq:app-5.10}
\end{align}

\begin{align}
\RS\bigl(\{\Xi_b^-\};\,\{\Sigma^+\};\,\{\pi^-\};\,\{K^-,\,\pi^-\};\,\Pzero\bigr){}+{}\RS\bigl(\{\Xi_b^-\};\,\{p\};\,\{K^-,\,\pi^-\};\,\{K^-\};\,\Pzero\bigr)&={}\,0 \tag{A5.11}\label{eq:app-5.11}
\end{align}

\begin{align}
\RS\bigl(\{\Lambda_b^0,\,\Xi_b^0\};\,\Bzero;\,\Pzero;\,\Pzero;\,\Pzero\bigr)&={}\,0 \tag{A5.12}\label{eq:app-5.12}
\end{align}

\begin{align}
\RS\bigl(\{\Lambda_b^0\};\,\Bzero;\,\{K^0\};\,\Pzero;\,\Pzero\bigr){}+{}\RS\bigl(\{\Xi_b^0\};\,\Bzero;\,\{\bar K^0\};\,\Pzero;\,\Pzero\bigr)&={}\,0 \tag{A5.13}\label{eq:app-5.13}
\end{align}

\begin{align}
&\phantom{{}+{}}\RS\bigl(\{\Lambda_b^0\};\,\Bzero;\,\{K^-,\,\pi^-\};\,\{K^+\};\,\Pzero\bigr){}+{}\RS\bigl(\{\Lambda_b^0\};\,\Bzero;\,\{\pi^-\};\,\{\pi^+\};\,\Pzero\bigr) \notag\\
&{}+{}\RS\bigl(\{\Xi_b^0\};\,\Bzero;\,\{K^-,\,\pi^-\};\,\{\pi^+\};\,\Pzero\bigr){}+{}\RS\bigl(\{\Xi_b^0\};\,\Bzero;\,\{K^+\};\,\{K^-\};\,\Pzero\bigr)={}\,0 \tag{A5.14}\label{eq:app-5.14}
\end{align}

\begin{align}
\RS\bigl(\{\Lambda_b^0,\,\Xi_b^0\};\,\Bzero;\,\{\bar K^0\};\,\{K^0\};\,\Pzero\bigr)&={}\,0 \tag{A5.15}\label{eq:app-5.15}
\end{align}

\begin{align}
&\phantom{{}+{}}\RS\bigl(\{\Lambda_b^0\};\,\Bzero;\,\{K^-,\,\pi^-\};\,\{\pi^+\};\,\{K^0\}\bigr){}+{}\RS\bigl(\{\Lambda_b^0\};\,\Bzero;\,\{K^0\};\,\{K^-\};\,\{K^+\}\bigr) \notag\\
&{}+{}\RS\bigl(\{\Lambda_b^0\};\,\Bzero;\,\{K^+\};\,\{\pi^-\};\,\{\bar K^0\}\bigr){}+{}\RS\bigl(\{\Xi_b^0\};\,\Bzero;\,\{K^-,\,\pi^-\};\,\{K^+\};\,\{\bar K^0\}\bigr) \notag\\
&{}+{}\RS\bigl(\{\Xi_b^0\};\,\Bzero;\,\{\pi^+\};\,\{K^0\};\,\{K^-\}\bigr){}+{}\RS\bigl(\{\Xi_b^0\};\,\Bzero;\,\{\pi^+\};\,\{\pi^-\};\,\{\bar K^0\}\bigr)={}\,0 \tag{A5.16}\label{eq:app-5.16}
\end{align}

\begin{align}
\RS\bigl(\{\Lambda_b^0\};\,\Bzero;\,\{\bar K^0\};\,\{K^0\};\,\{K^0\}\bigr){}+{}\RS\bigl(\{\Xi_b^0\};\,\Bzero;\,\{\bar K^0\};\,\{\bar K^0\};\,\{K^0\}\bigr)&={}\,0 \tag{A5.17}\label{eq:app-5.17}
\end{align}

\begin{align}
\RS\bigl(\{\Lambda_b^0\};\,\{n\};\,\Pzero;\,\Pzero;\,\Pzero\bigr){}+{}\RS\bigl(\{\Xi_b^0\};\,\{\Xi^0\};\,\Pzero;\,\Pzero;\,\Pzero\bigr)&={}\,0 \tag{A5.18}\label{eq:app-5.18}
\end{align}

\begin{align}
\RS\bigl(\{\Lambda_b^0,\,\Xi_b^0\};\,\{\Xi^0\};\,\{K^0\};\,\Pzero;\,\Pzero\bigr){}+{}\RS\bigl(\{\Lambda_b^0,\,\Xi_b^0\};\,\{n\};\,\{\bar K^0\};\,\Pzero;\,\Pzero\bigr)&={}\,0 \tag{A5.19}\label{eq:app-5.19}
\end{align}

\begin{align}
&\phantom{{}+{}}\RS\bigl(\{\Lambda_b^0\};\,\{\Xi^0\};\,\{\pi^-\};\,\{K^+\};\,\Pzero\bigr){}+{}\RS\bigl(\{\Lambda_b^0\};\,\{n\};\,\{K^-,\,\pi^-\};\,\{\pi^+\};\,\Pzero\bigr) \notag\\
&{}+{}\RS\bigl(\{\Lambda_b^0\};\,\{n\};\,\{K^+\};\,\{K^-\};\,\Pzero\bigr){}+{}\RS\bigl(\{\Xi_b^0\};\,\{\Xi^0\};\,\{K^+\};\,\{K^-,\,\pi^-\};\,\Pzero\bigr) \notag\\
&{}+{}\RS\bigl(\{\Xi_b^0\};\,\{\Xi^0\};\,\{\pi^-\};\,\{\pi^+\};\,\Pzero\bigr){}+{}\RS\bigl(\{\Xi_b^0\};\,\{n\};\,\{\pi^+\};\,\{K^-\};\,\Pzero\bigr)={}\,0 \tag{A5.20}\label{eq:app-5.20}
\end{align}

\begin{align}
&\phantom{{}+{}}\RS\bigl(\{\Lambda_b^0\};\,\{\Xi^0\};\,\{K^0\};\,\{K^0\};\,\Pzero\bigr){}+{}\RS\bigl(\{\Lambda_b^0\};\,\{n\};\,\{K^0\};\,\{\bar K^0\};\,\Pzero\bigr) \notag\\
&{}+{}\RS\bigl(\{\Xi_b^0\};\,\{\Xi^0\};\,\{K^0\};\,\{\bar K^0\};\,\Pzero\bigr){}+{}\RS\bigl(\{\Xi_b^0\};\,\{n\};\,\{\bar K^0\};\,\{\bar K^0\};\,\Pzero\bigr)={}\,0 \tag{A5.21}\label{eq:app-5.21}
\end{align}

\begin{align}
&\phantom{{}+{}}\RS\bigl(\{\Lambda_b^0\};\,\{\Sigma^-\};\,\{K^+,\,\pi^+\};\,\Pzero;\,\Pzero\bigr){}+{}\RS\bigl(\{\Lambda_b^0\};\,\{\Xi^-\};\,\{K^+\};\,\Pzero;\,\Pzero\bigr) \notag\\
&{}+{}\RS\bigl(\{\Xi_b^0\};\,\{\Sigma^-\};\,\{\pi^+\};\,\Pzero;\,\Pzero\bigr){}+{}\RS\bigl(\{\Xi_b^0\};\,\{\Xi^-\};\,\{K^+,\,\pi^+\};\,\Pzero;\,\Pzero\bigr)={}\,0 \tag{A5.22}\label{eq:app-5.22}
\end{align}

\begin{align}
&\phantom{{}+{}}\RS\bigl(\{\Lambda_b^0\};\,\{\Sigma^-\};\,\{K^+\};\,\{\bar K^0\};\,\Pzero\bigr){}+{}\RS\bigl(\{\Lambda_b^0\};\,\{\Sigma^-\};\,\{\pi^+\};\,\{K^0\};\,\Pzero\bigr) \notag\\
&{}+{}\RS\bigl(\{\Lambda_b^0\};\,\{\Xi^-\};\,\{K^0\};\,\{K^+,\,\pi^+\};\,\Pzero\bigr){}+{}\RS\bigl(\{\Xi_b^0\};\,\{\Sigma^-\};\,\{\bar K^0\};\,\{K^+,\,\pi^+\};\,\Pzero\bigr) \notag\\
&{}+{}\RS\bigl(\{\Xi_b^0\};\,\{\Xi^-\};\,\{K^+\};\,\{\bar K^0\};\,\Pzero\bigr){}+{}\RS\bigl(\{\Xi_b^0\};\,\{\Xi^-\};\,\{\pi^+\};\,\{K^0\};\,\Pzero\bigr)={}\,0 \tag{A5.23}\label{eq:app-5.23}
\end{align}

\begin{align}
&\phantom{{}+{}}\RS\bigl(\{\Lambda_b^0\};\,\{\Sigma^+\};\,\{\pi^-\};\,\Pzero;\,\Pzero\bigr){}+{}\RS\bigl(\{\Lambda_b^0\};\,\{p\};\,\{K^-,\,\pi^-\};\,\Pzero;\,\Pzero\bigr) \notag\\
&{}+{}\RS\bigl(\{\Xi_b^0\};\,\{\Sigma^+\};\,\{K^-,\,\pi^-\};\,\Pzero;\,\Pzero\bigr){}+{}\RS\bigl(\{\Xi_b^0\};\,\{p\};\,\{K^-\};\,\Pzero;\,\Pzero\bigr)={}\,0 \tag{A5.24}\label{eq:app-5.24}
\end{align}

\begin{align}
&\phantom{{}+{}}\RS\bigl(\{\Lambda_b^0\};\,\{\Sigma^+\};\,\{K^-,\,\pi^-\};\,\{K^0\};\,\Pzero\bigr){}+{}\RS\bigl(\{\Lambda_b^0\};\,\{p\};\,\{K^0\};\,\{K^-\};\,\Pzero\bigr) \notag\\
&{}+{}\RS\bigl(\{\Lambda_b^0\};\,\{p\};\,\{\pi^-\};\,\{\bar K^0\};\,\Pzero\bigr){}+{}\RS\bigl(\{\Xi_b^0\};\,\{\Sigma^+\};\,\{K^0\};\,\{K^-\};\,\Pzero\bigr) \notag\\
&{}+{}\RS\bigl(\{\Xi_b^0\};\,\{\Sigma^+\};\,\{\pi^-\};\,\{\bar K^0\};\,\Pzero\bigr){}+{}\RS\bigl(\{\Xi_b^0\};\,\{p\};\,\{K^-,\,\pi^-\};\,\{\bar K^0\};\,\Pzero\bigr)={}\,0 \tag{A5.25}\label{eq:app-5.25}
\end{align}

\begin{align}
\RS\bigl(\{\Xi_b^-\};\,\Bzero;\,\{K^{*-},\,\rho^-\};\,\Pzero;\,\Pzero\bigr)&={}\,0 \tag{A6.1}\label{eq:app-6.1}
\end{align}

\begin{align}
\RS\bigl(\{\Xi_b^-\};\,\Bzero;\,\Vzero;\,\{K^-,\,\pi^-\};\,\Pzero\bigr)&={}\,0 \tag{A6.2}\label{eq:app-6.2}
\end{align}

\begin{align}
\RS\bigl(\{\Xi_b^-\};\,\Bzero;\,\{K^{*0}\};\,\{K^-\};\,\Pzero\bigr){}+{}\RS\bigl(\{\Xi_b^-\};\,\Bzero;\,\{\bar K^{*0}\};\,\{\pi^-\};\,\Pzero\bigr)&={}\,0 \tag{A6.3}\label{eq:app-6.3}
\end{align}

\begin{align}
\RS\bigl(\{\Xi_b^-\};\,\Bzero;\,\{K^{*-}\};\,\{K^0\};\,\Pzero\bigr){}+{}\RS\bigl(\{\Xi_b^-\};\,\Bzero;\,\{\rho^-\};\,\{\bar K^0\};\,\Pzero\bigr)&={}\,0 \tag{A6.4}\label{eq:app-6.4}
\end{align}

{\color{red}
\begin{align}
\RS\bigl(\{\Xi_b^-\};\,\Bzero;\,\{K^{*+}\};\,\{K^-\};\,\{K^-\}\bigr){}+{}\RS\bigl(\{\Xi_b^-\};\,\Bzero;\,\{\rho^+\};\,\{\pi^-\};\,\{\pi^-\}\bigr)&={}\,0 \tag{A6.5}\label{eq:app-6.5}
\end{align}}

{\color{red}
\begin{align}
\RS\bigl(\{\Xi_b^-\};\,\Bzero;\,\{K^{*+},\,\rho^+\};\,\{\pi^-\};\,\{K^-\}\bigr)&={}\,0 \tag{A6.6}\label{eq:app-6.6}
\end{align}}

{\color{red}
\begin{align}
\RS\bigl(\{\Xi_b^-\};\,\Bzero;\,\{K^{*-}\};\,\{K^+\};\,\{K^-\}\bigr){}+{}\RS\bigl(\{\Xi_b^-\};\,\Bzero;\,\{\rho^-\};\,\{\pi^+\};\,\{\pi^-\}\bigr)&={}\,0 \tag{A6.7}\label{eq:app-6.7}
\end{align}}

{\color{red}
\begin{align}
\RS\bigl(\{\Xi_b^-\};\,\Bzero;\,\{K^{*-}\};\,\{\pi^-\};\,\{K^+\}\bigr){}+{}\RS\bigl(\{\Xi_b^-\};\,\Bzero;\,\{\rho^-\};\,\{\pi^+\};\,\{K^-\}\bigr)&={}\,0 \tag{A6.8}\label{eq:app-6.8}
\end{align}}

\begingroup
\color{red}
\begin{align}
\RS\bigl(\{\Xi_b^-\};\,\Bzero;\,\{K^{*-}\};\,\{\pi^+\};\,\{\pi^-\}\bigr){}+{}\RS\bigl(\{\Xi_b^-\};\,\Bzero;\,\{\rho^-\};\,\{K^+\};\,\{K^-\}\bigr)&={}\,0 \tag{A6.9}\label{eq:app-6.9}
\end{align}
\endgroup

\begin{align}
\RS\bigl(\{\Xi_b^-\};\,\Bzero;\,\Vzero;\,\{K^-\};\,\{K^0\}\bigr){}+{}\RS\bigl(\{\Xi_b^-\};\,\Bzero;\,\Vzero;\,\{\pi^-\};\,\{\bar K^0\}\bigr)&={}\,0 \tag{A6.10}\label{eq:app-6.10}
\end{align}

\begingroup
\color{red}
\begin{align}
\RS\bigl(\{\Xi_b^-\};\,\Bzero;\,\{\bar K^{*0}\};\,\{\pi^-\};\,\{K^0\}\bigr){}+{}\RS\bigl(\{\Xi_b^-\};\,\Bzero;\,\{K^{*0}\};\,\{K^-\};\,\{\bar K^0\}\bigr)&={}\,0 \tag{A6.11}\label{eq:app-6.11}
\end{align}
\endgroup

\begingroup
\color{red}
\begin{align}
\RS\bigl(\{\Xi_b^-\};\,\Bzero;\,\{\bar K^{*0}\};\,\{K^-\};\,\{K^0\}\bigr){}+{}\RS\bigl(\{\Xi_b^-\};\,\Bzero;\,\{K^{*0}\};\,\{\pi^-\};\,\{\bar K^0\}\bigr)&={}\,0 \tag{A6.12}\label{eq:app-6.12}
\end{align}
\endgroup

\begin{align}
\RS\bigl(\{\Xi_b^-\};\,\Bzero;\,\{K^{*-},\,\rho^-\};\,\{\bar K^0\};\,\{K^0\}\bigr)&={}\,0 \tag{A6.13}\label{eq:app-6.13}
\end{align}

\begin{align}
\RS\bigl(\{\Xi_b^-\};\,\{\Xi^0\};\,\{\rho^-\};\,\Pzero;\,\Pzero\bigr){}+{}\RS\bigl(\{\Xi_b^-\};\,\{n\};\,\{K^{*-}\};\,\Pzero;\,\Pzero\bigr)&={}\,0 \tag{A6.14}\label{eq:app-6.14}
\end{align}

\begin{align}
\RS\bigl(\{\Xi_b^-\};\,\{\Xi^0\};\,\Vzero;\,\{\pi^-\};\,\Pzero\bigr){}+{}\RS\bigl(\{\Xi_b^-\};\,\{n\};\,\Vzero;\,\{K^-\};\,\Pzero\bigr)&={}\,0 \tag{A6.15}\label{eq:app-6.15}
\end{align}

\begingroup
\color{red}
\begin{align}
\RS\bigl(\{\Xi_b^-\};\,\{\Xi^0\};\,\{K^{*0}\};\,\Pzero;\,\{K^-\}\bigr){}+{}\RS\bigl(\{\Xi_b^-\};\,\{n\};\,\{\bar K^{*0}\};\,\Pzero;\,\{\pi^-\}\bigr)&={}\,0 \tag{A6.16}\label{eq:app-6.16}
\end{align}
\endgroup

\begingroup
\color{red}
\begin{align}
\RS\bigl(\{\Xi_b^-\};\,\{\Xi^0\};\,\{K^{*0}\};\,\Pzero;\,\{\pi^-\}\bigr){}+{}\RS\bigl(\{\Xi_b^-\};\,\{n\};\,\{\bar K^{*0}\};\,\Pzero;\,\{K^-\}\bigr)&={}\,0 \tag{A6.17}\label{eq:app-6.17}
\end{align}
\endgroup

\begingroup
\color{red}
\begin{align}
\RS\bigl(\{\Xi_b^-\};\,\{\Xi^0\};\,\{K^{*-}\};\,\Pzero;\,\{K^0\}\bigr){}+{}\RS\bigl(\{\Xi_b^-\};\,\{n\};\,\{\rho^-\};\,\Pzero;\,\{\bar K^0\}\bigr)&={}\,0 \tag{A6.18}\label{eq:app-6.18}
\end{align}
\endgroup

\begingroup
\color{red}
\begin{align}
\RS\bigl(\{\Xi_b^-\};\,\{\Xi^0\};\,\{\rho^-\};\,\Pzero;\,\{K^0\}\bigr){}+{}\RS\bigl(\{\Xi_b^-\};\,\{n\};\,\{K^{*-}\};\,\Pzero;\,\{\bar K^0\}\bigr)&={}\,0 \tag{A6.19}\label{eq:app-6.19}
\end{align}
\endgroup

\begingroup
\color{red}
\begin{align}
\RS\bigl(\{\Xi_b^-\};\,\{\Xi^0\};\,\Vzero;\,\{K^-\};\,\{K^0\}\bigr){}+{}\RS\bigl(\{\Xi_b^-\};\,\{n\};\,\Vzero;\,\{\pi^-\};\,\{\bar K^0\}\bigr)&={}\,0 \tag{A6.20}\label{eq:app-6.20}
\end{align}
\endgroup

\begingroup
\color{red}
\begin{align}
\RS\bigl(\{\Xi_b^-\};\,\{\Xi^0\};\,\Vzero;\,\{\pi^-\};\,\{K^0\}\bigr){}+{}\RS\bigl(\{\Xi_b^-\};\,\{n\};\,\Vzero;\,\{K^-\};\,\{\bar K^0\}\bigr)&={}\,0 \tag{A6.21}\label{eq:app-6.21}
\end{align}
\endgroup

\begin{align}
\RS\bigl(\{\Xi_b^-\};\,\{\Sigma^-,\,\Xi^-\};\,\Vzero;\,\Pzero;\,\Pzero\bigr)&={}\,0 \tag{A6.22}\label{eq:app-6.22}
\end{align}

\begin{align}
\RS\bigl(\{\Xi_b^-\};\,\{\Sigma^-\};\,\{\bar K^{*0}\};\,\Pzero;\,\Pzero\bigr){}+{}\RS\bigl(\{\Xi_b^-\};\,\{\Xi^-\};\,\{K^{*0}\};\,\Pzero;\,\Pzero\bigr)&={}\,0 \tag{A6.23}\label{eq:app-6.23}
\end{align}

\begingroup
\color{red}
\begin{align}
\RS\bigl(\{\Xi_b^-\};\,\{\Sigma^-\};\,\{K^{*+}\};\,\Pzero;\,\{K^-\}\bigr){}+{}\RS\bigl(\{\Xi_b^-\};\,\{\Xi^-\};\,\{\rho^+\};\,\Pzero;\,\{\pi^-\}\bigr)&={}\,0 \tag{A6.24}\label{eq:app-6.24}
\end{align}
\endgroup

\begingroup
\color{red}
\begin{align}
\RS\bigl(\{\Xi_b^-\};\,\{\Sigma^-\};\,\{\rho^+\};\,\Pzero;\,\{K^-\}\bigr){}+{}\RS\bigl(\{\Xi_b^-\};\,\{\Xi^-\};\,\{K^{*+}\};\,\Pzero;\,\{\pi^-\}\bigr)&={}\,0 \tag{A6.25}\label{eq:app-6.25}
\end{align}
\endgroup

\begingroup
\color{red}
\begin{align}
\RS\bigl(\{\Xi_b^-\};\,\{\Sigma^-\};\,\{\rho^+\};\,\Pzero;\,\{\pi^-\}\bigr){}+{}\RS\bigl(\{\Xi_b^-\};\,\{\Xi^-\};\,\{K^{*+}\};\,\Pzero;\,\{K^-\}\bigr)&={}\,0 \tag{A6.26}\label{eq:app-6.26}
\end{align}
\endgroup

\begingroup
\color{red}
\begin{align}
\RS\bigl(\{\Xi_b^-\};\,\{\Sigma^-\};\,\{K^{*-}\};\,\Pzero;\,\{K^+\}\bigr){}+{}\RS\bigl(\{\Xi_b^-\};\,\{\Xi^-\};\,\{\rho^-\};\,\Pzero;\,\{\pi^+\}\bigr)&={}\,0 \tag{A6.27}\label{eq:app-6.27}
\end{align}
\endgroup

\begingroup
\color{red}
\begin{align}
\RS\bigl(\{\Xi_b^-\};\,\{\Sigma^-\};\,\{K^{*-}\};\,\Pzero;\,\{\pi^+\}\bigr){}+{}\RS\bigl(\{\Xi_b^-\};\,\{\Xi^-\};\,\{\rho^-\};\,\Pzero;\,\{K^+\}\bigr)&={}\,0 \tag{A6.28}\label{eq:app-6.28}
\end{align}
\endgroup

\begingroup
\color{red}
\begin{align}
\RS\bigl(\{\Xi_b^-\};\,\{\Sigma^-\};\,\{\rho^-\};\,\Pzero;\,\{\pi^+\}\bigr){}+{}\RS\bigl(\{\Xi_b^-\};\,\{\Xi^-\};\,\{K^{*-}\};\,\Pzero;\,\{K^+\}\bigr)&={}\,0 \tag{A6.29}\label{eq:app-6.29}
\end{align}
\endgroup

\begin{align}
\RS\bigl(\{\Xi_b^-\};\,\{\Sigma^-\};\,\Vzero;\,\{\bar K^0\};\,\Pzero\bigr){}+{}\RS\bigl(\{\Xi_b^-\};\,\{\Xi^-\};\,\Vzero;\,\{K^0\};\,\Pzero\bigr)&={}\,0 \tag{A6.30}\label{eq:app-6.30}
\end{align}

\begingroup
\color{red}
\begin{align}
\RS\bigl(\{\Xi_b^-\};\,\{\Sigma^-\};\,\{K^{*0}\};\,\Pzero;\,\{\bar K^0\}\bigr){}+{}\RS\bigl(\{\Xi_b^-\};\,\{\Xi^-\};\,\{\bar K^{*0}\};\,\Pzero;\,\{K^0\}\bigr)&={}\,0 \tag{A6.31}\label{eq:app-6.31}
\end{align}
\endgroup

\begingroup
\color{red}
\begin{align}
\RS\bigl(\{\Xi_b^-\};\,\{\Sigma^-\};\,\{\bar K^{*0}\};\,\Pzero;\,\{K^0\}\bigr){}+{}\RS\bigl(\{\Xi_b^-\};\,\{\Xi^-\};\,\{K^{*0}\};\,\Pzero;\,\{\bar K^0\}\bigr)&={}\,0 \tag{A6.32}\label{eq:app-6.32}
\end{align}
\endgroup

\begingroup
\color{red}
\begin{align}
\RS\bigl(\{\Xi_b^-\};\,\{\Sigma^-\};\,\Vzero;\,\{K^+\};\,\{K^-\}\bigr){}+{}\RS\bigl(\{\Xi_b^-\};\,\{\Xi^-\};\,\Vzero;\,\{\pi^+\};\,\{\pi^-\}\bigr)&={}\,0 \tag{A6.33}\label{eq:app-6.33}
\end{align}
\endgroup

\begingroup
\color{red}
\begin{align}
\RS\bigl(\{\Xi_b^-\};\,\{\Sigma^-\};\,\Vzero;\,\{\pi^+\};\,\{K^-\}\bigr){}+{}\RS\bigl(\{\Xi_b^-\};\,\{\Xi^-\};\,\Vzero;\,\{\pi^-\};\,\{K^+\}\bigr)&={}\,0 \tag{A6.34}\label{eq:app-6.34}
\end{align}
\endgroup

\begingroup
\color{red}
\begin{align}
\RS\bigl(\{\Xi_b^-\};\,\{\Sigma^-\};\,\Vzero;\,\{\pi^+\};\,\{\pi^-\}\bigr){}+{}\RS\bigl(\{\Xi_b^-\};\,\{\Xi^-\};\,\Vzero;\,\{K^+\};\,\{K^-\}\bigr)&={}\,0 \tag{A6.35}\label{eq:app-6.35}
\end{align}
\endgroup

\begin{align}
\RS\bigl(\{\Xi_b^-\};\,\{\Sigma^-,\,\Xi^-\};\,\Vzero;\,\{\bar K^0\};\,\{K^0\}\bigr)&={}\,0 \tag{A6.36}\label{eq:app-6.36}
\end{align}

\begingroup
\color{red}
\begin{align}
\RS\bigl(\{\Xi_b^-\};\,\{\Sigma^+\};\,\{K^{*-}\};\,\Pzero;\,\{\pi^-\}\bigr){}+{}\RS\bigl(\{\Xi_b^-\};\,\{p\};\,\{\rho^-\};\,\Pzero;\,\{K^-\}\bigr)&={}\,0 \tag{A6.37}\label{eq:app-6.37}
\end{align}
\endgroup

\begingroup
\color{red}
\begin{align}
\RS\bigl(\{\Xi_b^-\};\,\{\Sigma^+\};\,\{\rho^-\};\,\Pzero;\,\{K^-\}\bigr){}+{}\RS\bigl(\{\Xi_b^-\};\,\{p\};\,\{K^{*-}\};\,\Pzero;\,\{\pi^-\}\bigr)&={}\,0 \tag{A6.38}\label{eq:app-6.38}
\end{align}
\endgroup

\begingroup
\color{red}
\begin{align}
\RS\bigl(\{\Xi_b^-\};\,\{\Sigma^+\};\,\{\rho^-\};\,\Pzero;\,\{\pi^-\}\bigr){}+{}\RS\bigl(\{\Xi_b^-\};\,\{p\};\,\{K^{*-}\};\,\Pzero;\,\{K^-\}\bigr)&={}\,0 \tag{A6.39}\label{eq:app-6.39}
\end{align}
\endgroup

\begingroup
\color{red}
\begin{align}
\RS\bigl(\{\Xi_b^-\};\,\{\Sigma^+,\,p\};\,\Vzero;\,\{\pi^-\};\,\{K^-\}\bigr)&={}\,0 \tag{A6.40}\label{eq:app-6.40}
\end{align}
\endgroup

\begingroup
\color{red}
\begin{align}
\RS\bigl(\{\Xi_b^-\};\,\{\Sigma^+\};\,\Vzero;\,\{\pi^-\};\,\{\pi^-\}\bigr){}+{}\RS\bigl(\{\Xi_b^-\};\,\{p\};\,\Vzero;\,\{K^-\};\,\{K^-\}\bigr)&={}\,0 \tag{A6.41}\label{eq:app-6.41}
\end{align}
\endgroup

\begin{align}
\RS\bigl(\{\Lambda_b^0,\,\Xi_b^0\};\,\Bzero;\,\Vzero;\,\Pzero;\,\Pzero\bigr)&={}\,0 \tag{A6.42}\label{eq:app-6.42}
\end{align}

\begin{align}
\RS\bigl(\{\Lambda_b^0\};\,\Bzero;\,\{K^{*0}\};\,\Pzero;\,\Pzero\bigr){}+{}\RS\bigl(\{\Xi_b^0\};\,\Bzero;\,\{\bar K^{*0}\};\,\Pzero;\,\Pzero\bigr)&={}\,0 \tag{A6.43}\label{eq:app-6.43}
\end{align}

\begingroup
\color{red}
\begin{align}
\RS\bigl(\{\Lambda_b^0,\,\Xi_b^0\};\,\Bzero;\,\{K^{*+}\};\,\Pzero;\,\{K^-\}\bigr){}+{}\RS\bigl(\{\Lambda_b^0,\,\Xi_b^0\};\,\Bzero;\,\{\rho^+\};\,\Pzero;\,\{\pi^-\}\bigr)&={}\,0 \tag{A6.44}\label{eq:app-6.44}
\end{align}
\endgroup

\begingroup
\color{red}
\begin{align}
\RS\bigl(\{\Lambda_b^0\};\,\Bzero;\,\{K^{*+}\};\,\Pzero;\,\{\pi^-\}\bigr){}+{}\RS\bigl(\{\Xi_b^0\};\,\Bzero;\,\{\rho^+\};\,\Pzero;\,\{K^-\}\bigr)&={}\,0 \tag{A6.45}\label{eq:app-6.45}
\end{align}
\endgroup

\begingroup
\color{red}
\begin{align}
\RS\bigl(\{\Lambda_b^0,\,\Xi_b^0\};\,\Bzero;\,\{K^{*-}\};\,\Pzero;\,\{K^+\}\bigr){}+{}\RS\bigl(\{\Lambda_b^0,\,\Xi_b^0\};\,\Bzero;\,\{\rho^-\};\,\Pzero;\,\{\pi^+\}\bigr)&={}\,0 \tag{A6.46}\label{eq:app-6.46}
\end{align}
\endgroup

\begingroup
\color{red}
\begin{align}
\RS\bigl(\{\Lambda_b^0\};\,\Bzero;\,\{\rho^-\};\,\Pzero;\,\{K^+\}\bigr){}+{}\RS\bigl(\{\Xi_b^0\};\,\Bzero;\,\{K^{*-}\};\,\Pzero;\,\{\pi^+\}\bigr)&={}\,0 \tag{A6.47}\label{eq:app-6.47}
\end{align}
\endgroup

\begin{align}
\RS\bigl(\{\Lambda_b^0\};\,\Bzero;\,\Vzero;\,\{K^0\};\,\Pzero\bigr){}+{}\RS\bigl(\{\Xi_b^0\};\,\Bzero;\,\Vzero;\,\{\bar K^0\};\,\Pzero\bigr)&={}\,0 \tag{A6.48}\label{eq:app-6.48}
\end{align}

\begin{align}
\RS\bigl(\{\Lambda_b^0,\,\Xi_b^0\};\,\Bzero;\,\{\bar K^{*0}\};\,\{K^0\};\,\Pzero\bigr){}+{}\RS\bigl(\{\Lambda_b^0,\,\Xi_b^0\};\,\Bzero;\,\{K^{*0}\};\,\{\bar K^0\};\,\Pzero\bigr)&={}\,0 \tag{A6.49}\label{eq:app-6.49}
\end{align}

\begingroup
\color{red}
\begin{align}
\RS\bigl(\{\Lambda_b^0,\,\Xi_b^0\};\,\Bzero;\,\Vzero;\,\{K^+\};\,\{K^-\}\bigr){}+{}\RS\bigl(\{\Lambda_b^0,\,\Xi_b^0\};\,\Bzero;\,\Vzero;\,\{\pi^+\};\,\{\pi^-\}\bigr)&={}\,0 \tag{A6.50}\label{eq:app-6.50}
\end{align}
\endgroup

\begingroup
\color{red}
\begin{align}
\RS\bigl(\{\Lambda_b^0\};\,\Bzero;\,\Vzero;\,\{\pi^-\};\,\{K^+\}\bigr){}+{}\RS\bigl(\{\Xi_b^0\};\,\Bzero;\,\Vzero;\,\{\pi^+\};\,\{K^-\}\bigr)&={}\,0 \tag{A6.51}\label{eq:app-6.51}
\end{align}
\endgroup

\begingroup
\color{red}
\begin{align}
\RS\bigl(\{\Lambda_b^0\};\,\Bzero;\,\{K^{*0}\};\,\{K^+\};\,\{K^-\}\bigr){}+{}\RS\bigl(\{\Xi_b^0\};\,\Bzero;\,\{\bar K^{*0}\};\,\{\pi^+\};\,\{\pi^-\}\bigr)&={}\,0 \tag{A6.52}\label{eq:app-6.52}
\end{align}
\endgroup

\begingroup
\color{red}
\begin{align}
\RS\bigl(\{\Lambda_b^0,\,\Xi_b^0\};\,\Bzero;\,\{\bar K^{*0}\};\,\{\pi^-\};\,\{K^+\}\bigr){}+{}\RS\bigl(\{\Lambda_b^0,\,\Xi_b^0\};\,\Bzero;\,\{K^{*0}\};\,\{\pi^+\};\,\{K^-\}\bigr)&={}\,0 \tag{A6.53}\label{eq:app-6.53}
\end{align}
\endgroup

\begingroup
\color{red}
\begin{align}
\RS\bigl(\{\Lambda_b^0\};\,\Bzero;\,\{K^{*0}\};\,\{\pi^+\};\,\{\pi^-\}\bigr){}+{}\RS\bigl(\{\Xi_b^0\};\,\Bzero;\,\{\bar K^{*0}\};\,\{K^+\};\,\{K^-\}\bigr)&={}\,0 \tag{A6.54}\label{eq:app-6.54}
\end{align}
\endgroup

\begingroup
\color{red}
\begin{align}
\RS\bigl(\{\Lambda_b^0\};\,\Bzero;\,\{K^{*+}\};\,\{K^-\};\,\{K^0\}\bigr){}+{}\RS\bigl(\{\Xi_b^0\};\,\Bzero;\,\{\rho^+\};\,\{\pi^-\};\,\{\bar K^0\}\bigr)&={}\,0 \tag{A6.55}\label{eq:app-6.55}
\end{align}
\endgroup

\begingroup
\color{red}
\begin{align}
\RS\bigl(\{\Lambda_b^0\};\,\Bzero;\,\{\rho^+\};\,\{\pi^-\};\,\{K^0\}\bigr){}+{}\RS\bigl(\{\Xi_b^0\};\,\Bzero;\,\{K^{*+}\};\,\{K^-\};\,\{\bar K^0\}\bigr)&={}\,0 \tag{A6.56}\label{eq:app-6.56}
\end{align}
\endgroup

\begingroup
\color{red}
\begin{align}
\RS\bigl(\{\Lambda_b^0,\,\Xi_b^0\};\,\Bzero;\,\{K^{*+}\};\,\{\pi^-\};\,\{\bar K^0\}\bigr){}+{}\RS\bigl(\{\Lambda_b^0,\,\Xi_b^0\};\,\Bzero;\,\{\rho^+\};\,\{K^-\};\,\{K^0\}\bigr)&={}\,0 \tag{A6.57}\label{eq:app-6.57}
\end{align}
\endgroup

\begingroup
\color{red}
\begin{align}
\RS\bigl(\{\Lambda_b^0\};\,\Bzero;\,\{K^{*-}\};\,\{K^+\};\,\{K^0\}\bigr){}+{}\RS\bigl(\{\Xi_b^0\};\,\Bzero;\,\{\rho^-\};\,\{\pi^+\};\,\{\bar K^0\}\bigr)&={}\,0 \tag{A6.58}\label{eq:app-6.58}
\end{align}
\endgroup

\begingroup
\color{red}
\begin{align}
\RS\bigl(\{\Lambda_b^0\};\,\Bzero;\,\{\rho^-\};\,\{\pi^+\};\,\{K^0\}\bigr){}+{}\RS\bigl(\{\Xi_b^0\};\,\Bzero;\,\{K^{*-}\};\,\{K^+\};\,\{\bar K^0\}\bigr)&={}\,0 \tag{A6.59}\label{eq:app-6.59}
\end{align}
\endgroup

\begingroup
\color{red}
\begin{align}
\RS\bigl(\{\Lambda_b^0,\,\Xi_b^0\};\,\Bzero;\,\{K^{*-}\};\,\{\pi^+\};\,\{K^0\}\bigr){}+{}\RS\bigl(\{\Lambda_b^0,\,\Xi_b^0\};\,\Bzero;\,\{\rho^-\};\,\{K^+\};\,\{\bar K^0\}\bigr)&={}\,0 \tag{A6.60}\label{eq:app-6.60}
\end{align}
\endgroup

\begin{align}
\RS\bigl(\{\Lambda_b^0,\,\Xi_b^0\};\,\Bzero;\,\Vzero;\,\{\bar K^0\};\,\{K^0\}\bigr)&={}\,0 \tag{A6.61}\label{eq:app-6.61}
\end{align}

\begingroup
\color{red}
\begin{align}
\RS\bigl(\{\Lambda_b^0\};\,\Bzero;\,\{K^{*0}\};\,\{K^0\};\,\{\bar K^0\}\bigr){}+{}\RS\bigl(\{\Xi_b^0\};\,\Bzero;\,\{\bar K^{*0}\};\,\{K^0\};\,\{\bar K^0\}\bigr)&={}\,0 \tag{A6.62}\label{eq:app-6.62}
\end{align}
\endgroup

\begingroup
\color{red}
\begin{align}
\RS\bigl(\{\Lambda_b^0\};\,\Bzero;\,\{\bar K^{*0}\};\,\{K^0\};\,\{K^0\}\bigr){}+{}\RS\bigl(\{\Xi_b^0\};\,\Bzero;\,\{K^{*0}\};\,\{\bar K^0\};\,\{\bar K^0\}\bigr)&={}\,0 \tag{A6.63}\label{eq:app-6.63}
\end{align}
\endgroup

\begin{align}
\RS\bigl(\{\Lambda_b^0\};\,\{n\};\,\Vzero;\,\Pzero;\,\Pzero\bigr){}+{}\RS\bigl(\{\Xi_b^0\};\,\{\Xi^0\};\,\Vzero;\,\Pzero;\,\Pzero\bigr)&={}\,0 \tag{A6.64}\label{eq:app-6.64}
\end{align}

\begin{align}
\RS\bigl(\{\Lambda_b^0,\,\Xi_b^0\};\,\{\Xi^0\};\,\{K^{*0}\};\,\Pzero;\,\Pzero\bigr){}+{}\RS\bigl(\{\Lambda_b^0,\,\Xi_b^0\};\,\{n\};\,\{\bar K^{*0}\};\,\Pzero;\,\Pzero\bigr)&={}\,0 \tag{A6.65}\label{eq:app-6.65}
\end{align}

\begingroup
\color{red}
\begin{align}
\RS\bigl(\{\Lambda_b^0\};\,\{n\};\,\{\rho^+\};\,\Pzero;\,\{\pi^-\}\bigr){}+{}\RS\bigl(\{\Xi_b^0\};\,\{\Xi^0\};\,\{K^{*+}\};\,\Pzero;\,\{K^-\}\bigr)&={}\,0 \tag{A6.66}\label{eq:app-6.66}
\end{align}
\endgroup

\begingroup
\color{red}
\begin{align}
\RS\bigl(\{\Lambda_b^0,\,\Xi_b^0\};\,\{\Xi^0\};\,\{K^{*+}\};\,\Pzero;\,\{\pi^-\}\bigr){}+{}\RS\bigl(\{\Lambda_b^0,\,\Xi_b^0\};\,\{n\};\,\{\rho^+\};\,\Pzero;\,\{K^-\}\bigr)&={}\,0 \tag{A6.67}\label{eq:app-6.67}
\end{align}
\endgroup

\begingroup
\color{red}
\begin{align}
\RS\bigl(\{\Lambda_b^0\};\,\{n\};\,\{K^{*+}\};\,\Pzero;\,\{K^-\}\bigr){}+{}\RS\bigl(\{\Xi_b^0\};\,\{\Xi^0\};\,\{\rho^+\};\,\Pzero;\,\{\pi^-\}\bigr)&={}\,0 \tag{A6.68}\label{eq:app-6.68}
\end{align}
\endgroup

\begingroup
\color{red}
\begin{align}
\RS\bigl(\{\Lambda_b^0\};\,\{n\};\,\{\rho^-\};\,\Pzero;\,\{\pi^+\}\bigr){}+{}\RS\bigl(\{\Xi_b^0\};\,\{\Xi^0\};\,\{K^{*-}\};\,\Pzero;\,\{K^+\}\bigr)&={}\,0 \tag{A6.69}\label{eq:app-6.69}
\end{align}
\endgroup

\begingroup
\color{red}
\begin{align}
\RS\bigl(\{\Lambda_b^0,\,\Xi_b^0\};\,\{\Xi^0\};\,\{\rho^-\};\,\Pzero;\,\{K^+\}\bigr){}+{}\RS\bigl(\{\Lambda_b^0,\,\Xi_b^0\};\,\{n\};\,\{K^{*-}\};\,\Pzero;\,\{\pi^+\}\bigr)&={}\,0 \tag{A6.70}\label{eq:app-6.70}
\end{align}
\endgroup

\begingroup
\color{red}
\begin{align}
\RS\bigl(\{\Lambda_b^0\};\,\{n\};\,\{K^{*-}\};\,\Pzero;\,\{K^+\}\bigr){}+{}\RS\bigl(\{\Xi_b^0\};\,\{\Xi^0\};\,\{\rho^-\};\,\Pzero;\,\{\pi^+\}\bigr)&={}\,0 \tag{A6.71}\label{eq:app-6.71}
\end{align}
\endgroup

\begin{align}
\RS\bigl(\{\Lambda_b^0,\,\Xi_b^0\};\,\{\Xi^0\};\,\Vzero;\,\{K^0\};\,\Pzero\bigr){}+{}\RS\bigl(\{\Lambda_b^0,\,\Xi_b^0\};\,\{n\};\,\Vzero;\,\{\bar K^0\};\,\Pzero\bigr)&={}\,0 \tag{A6.72}\label{eq:app-6.72}
\end{align}

\begingroup
\color{red}
\begin{align}
\RS\bigl(\{\Lambda_b^0\};\,\{\Xi^0\};\,\{K^{*0}\};\,\Pzero;\,\{K^0\}\bigr){}+{}\RS\bigl(\{\Xi_b^0\};\,\{n\};\,\{\bar K^{*0}\};\,\Pzero;\,\{\bar K^0\}\bigr)&={}\,0 \tag{A6.73}\label{eq:app-6.73}
\end{align}
\endgroup

\begingroup
\color{red}
\begin{align}
\RS\bigl(\{\Lambda_b^0\};\,\{n\};\,\{\bar K^{*0}\};\,\Pzero;\,\{K^0\}\bigr){}+{}\RS\bigl(\{\Xi_b^0\};\,\{\Xi^0\};\,\{K^{*0}\};\,\Pzero;\,\{\bar K^0\}\bigr)&={}\,0 \tag{A6.74}\label{eq:app-6.74}
\end{align}
\endgroup

\begingroup
\color{red}
\begin{align}
\RS\bigl(\{\Lambda_b^0\};\,\{n\};\,\{K^{*0}\};\,\Pzero;\,\{\bar K^0\}\bigr){}+{}\RS\bigl(\{\Xi_b^0\};\,\{\Xi^0\};\,\{\bar K^{*0}\};\,\Pzero;\,\{K^0\}\bigr)&={}\,0 \tag{A6.75}\label{eq:app-6.75}
\end{align}
\endgroup

\begingroup
\color{red}
\begin{align}
\RS\bigl(\{\Lambda_b^0\};\,\{n\};\,\Vzero;\,\{\pi^+\};\,\{\pi^-\}\bigr){}+{}\RS\bigl(\{\Xi_b^0\};\,\{\Xi^0\};\,\Vzero;\,\{K^+\};\,\{K^-\}\bigr)&={}\,0 \tag{A6.76}\label{eq:app-6.76}
\end{align}
\endgroup

\begingroup
\color{red}
\begin{align}
\RS\bigl(\{\Lambda_b^0,\,\Xi_b^0\};\,\{\Xi^0\};\,\Vzero;\,\{\pi^-\};\,\{K^+\}\bigr){}+{}\RS\bigl(\{\Lambda_b^0,\,\Xi_b^0\};\,\{n\};\,\Vzero;\,\{\pi^+\};\,\{K^-\}\bigr)&={}\,0 \tag{A6.77}\label{eq:app-6.77}
\end{align}
\endgroup

\begingroup
\color{red}
\begin{align}
\RS\bigl(\{\Lambda_b^0\};\,\{n\};\,\Vzero;\,\{K^+\};\,\{K^-\}\bigr){}+{}\RS\bigl(\{\Xi_b^0\};\,\{\Xi^0\};\,\Vzero;\,\{\pi^+\};\,\{\pi^-\}\bigr)&={}\,0 \tag{A6.78}\label{eq:app-6.78}
\end{align}
\endgroup

\begingroup
\color{red}
\begin{align}
\RS\bigl(\{\Lambda_b^0\};\,\{\Xi^0\};\,\Vzero;\,\{K^0\};\,\{K^0\}\bigr){}+{}\RS\bigl(\{\Xi_b^0\};\,\{n\};\,\Vzero;\,\{\bar K^0\};\,\{\bar K^0\}\bigr)&={}\,0 \tag{A6.79}\label{eq:app-6.79}
\end{align}
\endgroup

\begingroup
\color{red}
\begin{align}
\RS\bigl(\{\Lambda_b^0\};\,\{n\};\,\Vzero;\,\{K^0\};\,\{\bar K^0\}\bigr){}+{}\RS\bigl(\{\Xi_b^0\};\,\{\Xi^0\};\,\Vzero;\,\{K^0\};\,\{\bar K^0\}\bigr)&={}\,0 \tag{A6.80}\label{eq:app-6.80}
\end{align}
\endgroup

\begingroup
\color{red}
\begin{align}
\RS\bigl(\{\Lambda_b^0\};\,\{\Sigma^-\};\,\{K^{*+}\};\,\Pzero;\,\Pzero\bigr){}+{}\RS\bigl(\{\Xi_b^0\};\,\{\Xi^-\};\,\{\rho^+\};\,\Pzero;\,\Pzero\bigr)&={}\,0 \tag{A6.81}\label{eq:app-6.81}
\end{align}
\endgroup

\begingroup
\color{red}
\begin{align}
\RS\bigl(\{\Lambda_b^0,\,\Xi_b^0\};\,\{\Sigma^-\};\,\{\rho^+\};\,\Pzero;\,\Pzero\bigr){}+{}\RS\bigl(\{\Lambda_b^0,\,\Xi_b^0\};\,\{\Xi^-\};\,\{K^{*+}\};\,\Pzero;\,\Pzero\bigr)&={}\,0 \tag{A6.82}\label{eq:app-6.82}
\end{align}
\endgroup

\begingroup
\color{red}
\begin{align}
\RS\bigl(\{\Lambda_b^0\};\,\{\Sigma^-\};\,\Vzero;\,\Pzero;\,\{K^+\}\bigr){}+{}\RS\bigl(\{\Xi_b^0\};\,\{\Xi^-\};\,\Vzero;\,\Pzero;\,\{\pi^+\}\bigr)&={}\,0 \tag{A6.83}\label{eq:app-6.83}
\end{align}
\endgroup

\begingroup
\color{red}
\begin{align}
\RS\bigl(\{\Lambda_b^0,\,\Xi_b^0\};\,\{\Sigma^-\};\,\Vzero;\,\Pzero;\,\{\pi^+\}\bigr){}+{}\RS\bigl(\{\Lambda_b^0,\,\Xi_b^0\};\,\{\Xi^-\};\,\Vzero;\,\Pzero;\,\{K^+\}\bigr)&={}\,0 \tag{A6.84}\label{eq:app-6.84}
\end{align}
\endgroup

\begingroup
\color{red}
\begin{align}
\RS\bigl(\{\Lambda_b^0\};\,\{\Sigma^-\};\,\{K^{*0}\};\,\Pzero;\,\{\pi^+\}\bigr){}+{}\RS\bigl(\{\Xi_b^0\};\,\{\Xi^-\};\,\{\bar K^{*0}\};\,\Pzero;\,\{K^+\}\bigr)&={}\,0 \tag{A6.85}\label{eq:app-6.85}
\end{align}
\endgroup

\begingroup
\color{red}
\begin{align}
\RS\bigl(\{\Lambda_b^0,\,\Xi_b^0\};\,\{\Sigma^-\};\,\{\bar K^{*0}\};\,\Pzero;\,\{K^+\}\bigr){}+{}\RS\bigl(\{\Lambda_b^0,\,\Xi_b^0\};\,\{\Xi^-\};\,\{K^{*0}\};\,\Pzero;\,\{\pi^+\}\bigr)&={}\,0 \tag{A6.86}\label{eq:app-6.86}
\end{align}
\endgroup

\begingroup
\color{red}
\begin{align}
\RS\bigl(\{\Lambda_b^0\};\,\{\Xi^-\};\,\{K^{*0}\};\,\Pzero;\,\{K^+\}\bigr){}+{}\RS\bigl(\{\Xi_b^0\};\,\{\Sigma^-\};\,\{\bar K^{*0}\};\,\Pzero;\,\{\pi^+\}\bigr)&={}\,0 \tag{A6.87}\label{eq:app-6.87}
\end{align}
\endgroup

\begingroup
\color{red}
\begin{align}
\RS\bigl(\{\Lambda_b^0\};\,\{\Sigma^-\};\,\{\rho^+\};\,\Pzero;\,\{K^0\}\bigr){}+{}\RS\bigl(\{\Xi_b^0\};\,\{\Xi^-\};\,\{K^{*+}\};\,\Pzero;\,\{\bar K^0\}\bigr)&={}\,0 \tag{A6.88}\label{eq:app-6.88}
\end{align}
\endgroup

\begingroup
\color{red}
\begin{align}
\RS\bigl(\{\Lambda_b^0\};\,\{\Xi^-\};\,\{K^{*+}\};\,\Pzero;\,\{K^0\}\bigr){}+{}\RS\bigl(\{\Xi_b^0\};\,\{\Sigma^-\};\,\{\rho^+\};\,\Pzero;\,\{\bar K^0\}\bigr)&={}\,0 \tag{A6.89}\label{eq:app-6.89}
\end{align}
\endgroup

\begingroup
\color{red}
\begin{align}
\RS\bigl(\{\Lambda_b^0,\,\Xi_b^0\};\,\{\Sigma^-\};\,\{K^{*+}\};\,\Pzero;\,\{\bar K^0\}\bigr){}+{}\RS\bigl(\{\Lambda_b^0,\,\Xi_b^0\};\,\{\Xi^-\};\,\{\rho^+\};\,\Pzero;\,\{K^0\}\bigr)&={}\,0 \tag{A6.90}\label{eq:app-6.90}
\end{align}
\endgroup

\begingroup
\color{red}
\begin{align}
\RS\bigl(\{\Lambda_b^0,\,\Xi_b^0\};\,\{\Sigma^-\};\,\Vzero;\,\{K^+\};\,\{\bar K^0\}\bigr){}+{}\RS\bigl(\{\Lambda_b^0,\,\Xi_b^0\};\,\{\Xi^-\};\,\Vzero;\,\{\pi^+\};\,\{K^0\}\bigr)&={}\,0 \tag{A6.91}\label{eq:app-6.91}
\end{align}
\endgroup

\begingroup
\color{red}
\begin{align}
\RS\bigl(\{\Lambda_b^0\};\,\{\Sigma^-\};\,\Vzero;\,\{\pi^+\};\,\{K^0\}\bigr){}+{}\RS\bigl(\{\Xi_b^0\};\,\{\Xi^-\};\,\Vzero;\,\{K^+\};\,\{\bar K^0\}\bigr)&={}\,0 \tag{A6.92}\label{eq:app-6.92}
\end{align}
\endgroup

\begingroup
\color{red}
\begin{align}
\RS\bigl(\{\Lambda_b^0\};\,\{\Xi^-\};\,\Vzero;\,\{K^+\};\,\{K^0\}\bigr){}+{}\RS\bigl(\{\Xi_b^0\};\,\{\Sigma^-\};\,\Vzero;\,\{\pi^+\};\,\{\bar K^0\}\bigr)&={}\,0 \tag{A6.93}\label{eq:app-6.93}
\end{align}
\endgroup

\begingroup
\color{red}
\begin{align}
\RS\bigl(\{\Lambda_b^0\};\,\{p\};\,\{\rho^-\};\,\Pzero;\,\Pzero\bigr){}+{}\RS\bigl(\{\Xi_b^0\};\,\{\Sigma^+\};\,\{K^{*-}\};\,\Pzero;\,\Pzero\bigr)&={}\,0 \tag{A6.94}\label{eq:app-6.94}
\end{align}
\endgroup

\begingroup
\color{red}
\begin{align}
\RS\bigl(\{\Lambda_b^0,\,\Xi_b^0\};\,\{\Sigma^+\};\,\{\rho^-\};\,\Pzero;\,\Pzero\bigr){}+{}\RS\bigl(\{\Lambda_b^0,\,\Xi_b^0\};\,\{p\};\,\{K^{*-}\};\,\Pzero;\,\Pzero\bigr)&={}\,0 \tag{A6.95}\label{eq:app-6.95}
\end{align}
\endgroup

\begingroup
\color{red}
\begin{align}
\RS\bigl(\{\Lambda_b^0\};\,\{p\};\,\Vzero;\,\Pzero;\,\{\pi^-\}\bigr){}+{}\RS\bigl(\{\Xi_b^0\};\,\{\Sigma^+\};\,\Vzero;\,\Pzero;\,\{K^-\}\bigr)&={}\,0 \tag{A6.96}\label{eq:app-6.96}
\end{align}
\endgroup

\begingroup
\color{red}
\begin{align}
\RS\bigl(\{\Lambda_b^0,\,\Xi_b^0\};\,\{\Sigma^+\};\,\Vzero;\,\Pzero;\,\{\pi^-\}\bigr){}+{}\RS\bigl(\{\Lambda_b^0,\,\Xi_b^0\};\,\{p\};\,\Vzero;\,\Pzero;\,\{K^-\}\bigr)&={}\,0 \tag{A6.97}\label{eq:app-6.97}
\end{align}
\endgroup

\begingroup
\color{red}
\begin{align}
\RS\bigl(\{\Lambda_b^0,\,\Xi_b^0\};\,\{\Sigma^+\};\,\{K^{*0}\};\,\Pzero;\,\{K^-\}\bigr){}+{}\RS\bigl(\{\Lambda_b^0,\,\Xi_b^0\};\,\{p\};\,\{\bar K^{*0}\};\,\Pzero;\,\{\pi^-\}\bigr)&={}\,0 \tag{A6.98}\label{eq:app-6.98}
\end{align}
\endgroup

\begingroup
\color{red}
\begin{align}
\RS\bigl(\{\Lambda_b^0\};\,\{\Sigma^+\};\,\{K^{*0}\};\,\Pzero;\,\{\pi^-\}\bigr){}+{}\RS\bigl(\{\Xi_b^0\};\,\{p\};\,\{\bar K^{*0}\};\,\Pzero;\,\{K^-\}\bigr)&={}\,0 \tag{A6.99}\label{eq:app-6.99}
\end{align}
\endgroup

\begingroup
\color{red}
\begin{align}
\RS\bigl(\{\Lambda_b^0\};\,\{p\};\,\{K^{*0}\};\,\Pzero;\,\{K^-\}\bigr){}+{}\RS\bigl(\{\Xi_b^0\};\,\{\Sigma^+\};\,\{\bar K^{*0}\};\,\Pzero;\,\{\pi^-\}\bigr)&={}\,0 \tag{A6.100}\label{eq:app-6.100}
\end{align}
\endgroup

\begingroup
\color{red}
\begin{align}
\RS\bigl(\{\Lambda_b^0,\,\Xi_b^0\};\,\{\Sigma^+\};\,\{K^{*-}\};\,\Pzero;\,\{K^0\}\bigr){}+{}\RS\bigl(\{\Lambda_b^0,\,\Xi_b^0\};\,\{p\};\,\{\rho^-\};\,\Pzero;\,\{\bar K^0\}\bigr)&={}\,0 \tag{A6.101}\label{eq:app-6.101}
\end{align}
\endgroup

\begingroup
\color{red}
\begin{align}
\RS\bigl(\{\Lambda_b^0\};\,\{\Sigma^+\};\,\{\rho^-\};\,\Pzero;\,\{K^0\}\bigr){}+{}\RS\bigl(\{\Xi_b^0\};\,\{p\};\,\{K^{*-}\};\,\Pzero;\,\{\bar K^0\}\bigr)&={}\,0 \tag{A6.102}\label{eq:app-6.102}
\end{align}
\endgroup

\begingroup
\color{red}
\begin{align}
\RS\bigl(\{\Lambda_b^0\};\,\{p\};\,\{K^{*-}\};\,\Pzero;\,\{K^0\}\bigr){}+{}\RS\bigl(\{\Xi_b^0\};\,\{\Sigma^+\};\,\{\rho^-\};\,\Pzero;\,\{\bar K^0\}\bigr)&={}\,0 \tag{A6.103}\label{eq:app-6.103}
\end{align}
\endgroup

\begingroup
\color{red}
\begin{align}
\RS\bigl(\{\Lambda_b^0,\,\Xi_b^0\};\,\{\Sigma^+\};\,\Vzero;\,\{K^-\};\,\{K^0\}\bigr){}+{}\RS\bigl(\{\Lambda_b^0,\,\Xi_b^0\};\,\{p\};\,\Vzero;\,\{\pi^-\};\,\{\bar K^0\}\bigr)&={}\,0 \tag{A6.104}\label{eq:app-6.104}
\end{align}
\endgroup

\begingroup
\color{red}
\begin{align}
\RS\bigl(\{\Lambda_b^0\};\,\{\Sigma^+\};\,\Vzero;\,\{\pi^-\};\,\{K^0\}\bigr){}+{}\RS\bigl(\{\Xi_b^0\};\,\{p\};\,\Vzero;\,\{K^-\};\,\{\bar K^0\}\bigr)&={}\,0 \tag{A6.105}\label{eq:app-6.105}
\end{align}
\endgroup

\begingroup
\color{red}
\begin{align}
\RS\bigl(\{\Lambda_b^0\};\,\{p\};\,\Vzero;\,\{K^-\};\,\{K^0\}\bigr){}+{}\RS\bigl(\{\Xi_b^0\};\,\{\Sigma^+\};\,\Vzero;\,\{\pi^-\};\,\{\bar K^0\}\bigr)&={}\,0 \tag{A6.106}\label{eq:app-6.106}
\end{align}
\endgroup

\begin{align}
\RS\bigl(\{\Lambda_c^+\};\,\Bzero;\,\{K^+\}\bigr){}+{}\RS\bigl(\{\Xi_c^+\};\,\Bzero;\,\{\pi^+\}\bigr)&={}\,0 \tag{A7.1}\label{eq:app-7.1}
\end{align}

\begin{align}
\RS\bigl(\{\Lambda_c^+\};\,\{p\};\,\Pzero\bigr){}+{}\RS\bigl(\{\Xi_c^+\};\,\{\Sigma^+\};\,\Pzero\bigr)&={}\,0 \tag{A7.2}\label{eq:app-7.2}
\end{align}

\begin{align}
\RS\bigl(\{\Xi_c^0\};\,\Bzero;\,\Pzero\bigr)&={}\,0 \tag{A7.3}\label{eq:app-7.3}
\end{align}

\begin{align}
\RS\bigl(\{\Lambda_c^+\};\,\Bzero;\,\{K^{*+}\}\bigr){}+{}\RS\bigl(\{\Xi_c^+\};\,\Bzero;\,\{\rho^+\}\bigr)&={}\,0 \tag{A8.1}\label{eq:app-8.1}
\end{align}

\begin{align}
\RS\bigl(\{\Lambda_c^+\};\,\{p\};\,\Vzero\bigr){}+{}\RS\bigl(\{\Xi_c^+\};\,\{\Sigma^+\};\,\Vzero\bigr)&={}\,0 \tag{A8.2}\label{eq:app-8.2}
\end{align}

\begin{align}
\RS\bigl(\{\Xi_c^0\};\,\Bzero;\,\Vzero\bigr)&={}\,0 \tag{A8.3}\label{eq:app-8.3}
\end{align}

\begin{align}
\RS\bigl(\{\Lambda_c^+\};\,\{\Delta^+\};\,\Pzero\bigr){}+{}\RS\bigl(\{\Xi_c^+\};\,\{\Sigma^{*+}\};\,\Pzero\bigr)&={}\,0 \tag{A9.1}\label{eq:app-9.1}
\end{align}

\begin{align}
\RS\bigl(\{\Xi_c^0\};\,\{\Sigma^{*0}\};\,\Pzero\bigr)&={}\,0 \tag{A9.2}\label{eq:app-9.2}
\end{align}

\begin{align}
\RS\bigl(\{\Xi_b^-\};\,\Bzero;\,\{K^-,\,\pi^-\}\bigr)&={}\,0 \tag{A10.1}\label{eq:app-10.1}
\end{align}

\begin{align}
\RS\bigl(\{\Xi_b^-\};\,\{\Sigma^-,\,\Xi^-\};\,\Pzero\bigr)&={}\,0 \tag{A10.2}\label{eq:app-10.2}
\end{align}

\begin{align}
\RS\bigl(\{\Lambda_b^0,\,\Xi_b^0\};\,\Bzero;\,\Pzero\bigr)&={}\,0 \tag{A10.3}\label{eq:app-10.3}
\end{align}

\begin{align}
\RS\bigl(\{\Lambda_b^0\};\,\Bzero;\,\{K^0\}\bigr){}+{}\RS\bigl(\{\Xi_b^0\};\,\Bzero;\,\{\bar K^0\}\bigr)&={}\,0 \tag{A10.4}\label{eq:app-10.4}
\end{align}

\begin{align}
\RS\bigl(\{\Lambda_b^0\};\,\{n\};\,\Pzero\bigr){}+{}\RS\bigl(\{\Xi_b^0\};\,\{\Xi^0\};\,\Pzero\bigr)&={}\,0 \tag{A10.5}\label{eq:app-10.5}
\end{align}

\begin{align}
\RS\bigl(\{\Xi_b^-\};\,\Bzero;\,\{K^{*-},\,\rho^-\}\bigr)&={}\,0 \tag{A11.1}\label{eq:app-11.1}
\end{align}

\begin{align}
\RS\bigl(\{\Xi_b^-\};\,\{\Sigma^-,\,\Xi^-\};\,\Vzero\bigr)&={}\,0 \tag{A11.2}\label{eq:app-11.2}
\end{align}

\begin{align}
\RS\bigl(\{\Lambda_b^0,\,\Xi_b^0\};\,\Bzero;\,\Vzero\bigr)&={}\,0 \tag{A11.3}\label{eq:app-11.3}
\end{align}

\begin{align}
\RS\bigl(\{\Lambda_b^0\};\,\Bzero;\,\{K^{*0}\}\bigr){}+{}\RS\bigl(\{\Xi_b^0\};\,\Bzero;\,\{\bar K^{*0}\}\bigr)&={}\,0 \tag{A11.4}\label{eq:app-11.4}
\end{align}

\begin{align}
\RS\bigl(\{\Lambda_b^0\};\,\{n\};\,\Vzero\bigr){}+{}\RS\bigl(\{\Xi_b^0\};\,\{\Xi^0\};\,\Vzero\bigr)&={}\,0 \tag{A11.5}\label{eq:app-11.5}
\end{align}

\begin{align}
\RS\bigl(\{\Xi_b^-\};\,\{\Sigma^{*-},\,\Xi^{*-}\};\,\Pzero\bigr)&={}\,0 \tag{A12.1}\label{eq:app-12.1}
\end{align}

\begin{align}
\RS\bigl(\{\Lambda_b^0\};\,\{\Delta^0\};\,\Pzero\bigr){}+{}\RS\bigl(\{\Xi_b^0\};\,\{\Xi^{*0}\};\,\Pzero\bigr)&={}\,0 \tag{A12.2}\label{eq:app-12.2}
\end{align}

\begin{align}
\RS\bigl(\{\Lambda_b^0,\,\Xi_b^0\};\,\{\Sigma^{*0}\};\,\Pzero\bigr)&={}\,0 \tag{A12.3}\label{eq:app-12.3}
\end{align}

}
% END INLINE APPENDIX

% BEGIN INLINE REFERENCES

% END INLINE REFERENCES

\end{document}